\documentclass[preprint,showpacs]{revtex4}

\usepackage{graphicx, latexsym}
\usepackage{bm}
\usepackage{color}
\usepackage{amsmath}
\usepackage{natbib}
\usepackage{graphicx}				 
\usepackage{array,makecell}

\begin{document}

\title{A new and alternative look at NonLinear Alfv\'enic States}
\author{Swadesh M. Mahajan}
\email{mahajan@mail.utexas.edu}
\affiliation{Institute for Fusion Studies, The University of Texas at Austin, Texas 78712, USA.}
\date{\today}

\begin{abstract}
A new formalism for the nonlinear Alfv\'enic states sustainable in Hall Magnetohydrodynamics is developed in a complete basis provided by the circularly polarized Beltrami Vectors, the eigenstates of linear HMHD. Nonlinear HMHD is, then, reduced to a rather simple looking set of scalar equations from which a model problem of three interacting Beltrami modes is formulated and analytically solved. The triplet interactions span a variety of familiar nonlinear processes leading to a redistribution as well as periodic exchange of energy. The energy exchange processes (whose strength is measured by an energy exchange /depletion time) will, perhaps, play a dominant role in determining the spectral content of an eventual Alfv\'enic state. All nonlinearities (sensitive functions of the interacting wave vectors) operate at par, and none is dominant over any substantial region of k-space; their intricate interplay prevents a ``universal" picture from emerging; few generalizations on the processes that may, for instance, lead to a turbulent state, are possible. However, the theory can definitely claim: 1) the energy tends to flow from lower to higher $k$, and 2) the higher $k_z$ (in the direction of the ambient magnetic field) components of a mode with a given $k$ are depleted/oscillate faster -in some cases much faster. It is noteworthy that the mode coupling is the strongest (with the shortest depletion time) when the participating wave vectors are nearly perpendicular; perhaps, an expected consequence of the curl (cross product) nonlinearities.  Numerical simulations will be necessary to help create a fully reliable picture.
\end{abstract}

\pacs{04.70.Bw, 52.27.Ny, 52.35.We, 95.30.Qd}


\maketitle


\section{Introduction} 

Nonlinear  Alfv\'enic  states, both coherent and turbulent, have been so thoroughly investigated, especially in the context of magnetohydrodynamics (MHD)~\cite{IR63, Kra65, SMM83, Hi84, SG94, SG95, NB97, CV00, GNNP00, MG01, CZV02,LPS02, MDG03, MG05, OM05, SC05, BL06, G07, BL08, K65, ZLF 92}, that I undertake this task of presenting a ``new formulation" with great trepidation. However, in light of the many advances in the theory and simulations of  Alfv\'enic turbulence \citep{Bis03,ZMD04,MV11,Dav,Gal18,Sch20, GP04,Fre14} as well as the accompanying data from laboratory and space plasmas - with the solar wind constituting a classic example \citep{ACS13,BC16,SHH20,OE21} - such an endeavor is arguably timely.

When one allows our vast experience with extended MHD, to wit, systems containing physics beyond pure MHD, to dictate future directions, one must confront the fact that a much more natural description of a general elementary Alfv\'enic mode is provided by a circularly polarized wave and not a Fourier mode with arbitrary polarization. The most widely used representation for a circularly polarized wave in Alfv\'enic systems is probably the Beltrami vector function (a summary follows); being an eigenstate of the curl operator, it has the additional crucial advantage that it tends to simplify the complicated "curl -dominated" nonlinearities in extended MHD models \cite{SI,MY98,ML1}.

Thus, one of the main motivation for this effort is the expectation that when we attempt to understand the Alfv\'enic state in terms of its ``normal modes" as described above, the eventual processed system will be rendered simpler and more transparent and, therefore, easier to explore by the standard tools. Numerical simulations will, perhaps, be less time consuming and one could also hope to extract crucial information by straightforward analysis that may not be accessible in a more complicated formalism. The Beltrami formalism automatically serves to constrain the system to the interaction of the allowed modes of oscillations.

I believe that this paper makes a preliminary case that both these expectations are, indeed, legitimate: the system of equations, derived with the expectation of that they will be numerically simulated, appear to be considerably simpler. However the numerical advantages (if any) of this formalism will need to be demonstrated by actual simulations, and constitutes a topic for future collaborative work. The simulation of the extended MHD system is, thus, not the objective of this paper. 

There are two easily distinguishable but connected parts of the text: In Part I, I will derive the basic set of Beltrami-transformed HMHD equations, discuss very general features of the nonlinear interaction/ coupling coefficients, and develop  an even simpler set by exploiting some  results from the very highly studied linear approximation. Part II, will be, more or less, fully devoted  to analytical investigations into the complex nature of the Alfv\'enic nonlinearities;  it will be done by working out the evolution of a truncated but complete set of a  limited number of resonantly interacting modes. A natural choice for a system with quadratic nonlinearities is the three mode interaction; it will be studied in considerable detail. Although the three mode interaction span a broad class of nonlinear processes- with and without energy exchange- the mode envelopes evolve at a definite characteristic time (time scales much slower than the linear Alfv\'enic times). The energy exchanges could be either periodic or unidirectional.  Calculating the speed and  direction (in the wave vector space) of these exchanges constitutes a major goal.  This model calculation lays no claim  to capturing all the essentials of a turbulent state, but will hopefully give glimpses of the processes that will lead to one. Needless to state that the word "mode" in this paper refers to a Beltrami mode - a vector Fourier mode whose polarization is fully determined by the wave vector. 

The most important finding of this initial analytical effort is that the action of the nonlinear interactions is quite variegated, and defies a facile ``universal" description. It reveals, at the same time, some expected and some not-so-obvious processes that build up a nonlinear Alfv\'enic state. For relative simplicity, I have concentrated on a restricted but highly relevant region 
of $k$-space in which $k_z<< k, k_{\perp}$ where $k_z$ and $k_{\perp}$ are the projection of the wave vector in the directions parallel and perpendicular to the ambient magnetic field, respectively, whereas $k$ is the modulus of the wave vector. In actuality, though, this choice is not necessary because the formalism is completely amenable to exploring $k_z\simeq k_{\perp}$ part of the spectrum.

The study will span both branches of the linear Alfv\'enic oscillations - the Alfv\'en-Whistler (AW $\equiv +$) and the Alfv\'en-Cyclotron (AC$\equiv -$) waves are described in Sec.VA. A more complete understanding of the Alfv\'enic state, therefore, must comprise intra-branch as well as inter-branch interactions. It so happens that the solvability conditions for the particular model problem of resonant energy exchange between two modes, are somewhat hard to satisfy for the inter-branch processes. But this should be seen as an exception; generally  the interaction of two different branches of oscillations would significantly contribute to  Alfv\'enic turbulence.  

Of all the possible interactions, I will dwell on the resonant interactions which are expected to be most responsible for energy exchange between the waves. The time scales of these resonant processes, signifying the time scales on which the wave envelopes evolve,  must be assumed to be much slower than the Alfv\'enic oscillation times; it is this time hierarchy that guarantees analytic solvability of the model. We will see that the the resonant condition imposes a relationship between $ k_{1z}$ and $k_{2z}$ that further restricts the $k$ domain that can contribute to the resonant process. 
  
Due to the complexity and intricacy of the calculations, the salient findings of the model are adumbrated below:

\begin{itemize} 
\item The first realization is that the modes with exactly parallel (antiparallel) wave numbers do not interact at all; the interaction kernel $I_{mn}$, that captures the vectorial essence of the nonlinearities, is identically zero independent of any details including the lengths of the interacting wave vectors. The implication, of course, is that a system of Alfv\'en waves, constrained to parallel wave vectors, must behave as linear waves even when they have arbitrary amplitudes. It is not that the nonlinear terms are neglected; they just make no contributions to the dynamics. This feature of the system, in some form, has been known since a long time ~\cite{W1,SV05}. It was, however, exploited in prior publications to show the linear superposition of what could be called Linear-Nonlinear waves, i.e, nonlinear waves with a linear characteristics including the dispersion relation ~\cite{MM09,XZ15,ALM16,LB16,AY16}.

\item  It turns out that $I_{mn}$ is mostly controlled by the relative orientation and magnitudes of the participating ${\bf k}_{1{\perp}}, {\bf k}_{2{\perp}}$; their relative orientation controls the interaction strength qualitatively while their relative magnitudes have  mostly quantitative effects. The interaction kernel has such strong variation over k space that detailed calculations are necessary. in fact, such calculations form the bulk of Part II.

\item A very important feature of the Nonlinear HMHD is that all nonlinearities are well matched, to wit, there is no dominant nonlinearity in any large enough $k$ domain. One would expect that in the MHD regime of low $k$, the ${\bf v}\times {\bf b}$ nonlinearity should dominate others [namely ${\bf b}\times (\nabla\times{\bf b})$ and ${\bf v}\times (\nabla\times{\bf v})$] that have an extra gradient (corresponding to an extra $k$ factor). However, it is not so because its leading order manifestation (order unity) gets cancelled. One can see  from  Eqs.[\ref{Mu1}- \ref{Muh}] that the factors $\mu_{1, 2, h}$ (which, along with $I_{12}$, determine the total strength of the interaction) end up being of order $k$ (same order as the contributions, for instance, of the Hall nonlinearity) due to an intricate mixing of all contributing nonlinearities.

\item  This cancellation pertains to all cases when the component modes are within the same branch $ (+ ++)$ or $( - - -)$. The only exception would have been  the inter-branch coupling (two modes on branch and the third on the other) of near counter propagating modes;  the ${\bf v}\times {\bf b}$ nonlinearity would, then, manifest itself strongly to order unity. However, for this class pf coupling, it is hard to satisfy the conditions necessary for the resonant interaction to lead to a meaningful initial value problem. This process is , thus, not investigated in the model calculation worked out here.

\item The overall strength/efficacy of the nonlinear interaction for the model problem involving two sets of modes is measured in terms of what we label the energy exchange time $t_{exchange}$ - it is the time needed for substantial energy exchange (periodic or directional) between modes.  In fact, $t_{exchange}$ may be recognized as the fundamental signature of nonlinear HMHD. 

\item The interaction coefficients (and eventually the $t_{exchange}$) show great sensitivity to the relative orientation and magnitude of [${\bf k_{1\perp}}, {\bf k_{2\perp}}$]. One must be rather careful in their evaluation. The calculations are rather straightforward but \emph{a priori} predictions, based on assuming a dominant nonlinearity or process, for instance, may not be accurate or even qualitatively correct. This realization is of fundamental importance to the further development of the subject.

\item  A whole range of nonlinear processes, with directional as well as periodic energy exchange, take place as do processes where there is no effective energy exchange within the participating modes. Despite the sensitivity of the interaction to its location in $k$, there are a few general characteristic of the energy exchange process that come nearest to being universal: 1) In all processes that have directional energy exchange, it is the highest wave number mode to which most of the initial energy is eventually shifted, 2) At least in some regions of the k-space, the energy exchange is faster amongst two modes with different $k_z$ but approximately the same $k$ (possible when $k_z << k$), the one with larger $k_z$ has faster energy depletion time. These two processes could lead to the following scenario: If one were to wait long enough, such an Alfv\'enic system will get richer in both low $k_z/k$ and high $ k$ part of the spectrum. In other words, the asymptotic spectrum will be characterized by $k_\perp>>k_z$.  The scaling of  $t_{depletion}$ with $k_z$ varies a great deal within the explored range of the k-space. The strongest $k_z$ variation is seen in the $\pi$ sector where  $t_{exchange}\simeq k_{1z}^{-9/2}$.  

I will defer to the last Section the difficult job of putting this work in perspective; the reader must  get familiar with the calculation  before I attempt a critical examination of both the formalism and the model calculation.

\section{Beltrami vectors and Beltrami transforms}

We begin with a short introduction to Beltrami vectors. A Beltrami vector
(with an index n) is defined as (I am using a slightly unusual notation)
\begin{equation}\label{BV}
{\bf Q}_{n} = \hat{e_{n}} e^{i {\bf {k}}_n \cdot {\bf x}}	 
\end{equation}
where the wave vector 
\begin{equation}\label{WaveV}
{\bf k_{n}} = k_n \hat{k_n}
\end{equation}
is written in terms of the positive definite magnitude $k_n$, and the unit vector $\hat{k_n}= -\hat{k_{-n}}$; the latter
changes sign with the index $n$ . The complex polarization vector (indicating circular polarization)
\begin{equation}\label{PolVPre}
\hat{e_{n}} \propto(\hat{e}_{n1} + i \hat{e}_{n2})
\end{equation}
will be  so constructed that  $\hat{e}_{n1}$, $\hat{e}_{n2}$, and $\hat{k_n}$ form an orthogonal right-handed triad of unit vectors. 
Anticipating that we will place the ambient magnetic field along the z direction ( ${\bf B}_0= B_0 \hat{e_{z}}$), one may readily
use the representation 
\begin{equation} \label{Rep PolV}
\hat{e}_{n1} = \frac{k_n}{k_{n\perp}\sqrt{2}}(\hat{e}_z \times \hat{k_n}), \quad  \quad  \hat{e}_{n2}=  \frac{k_n}{k_{n\perp}\sqrt{2}}(\hat{k_n} \times (\hat{e}_z \times \hat{k_n})),
\end{equation} 
with the required normalizations to insure the orthonormality
\begin{equation}\label{PropPolV}
\hat{e_{n}}\cdot\hat{e_{n}}=0, \quad   \hat{e_{n}}\cdot {\hat{e_{n}}}^*=1
\end{equation}
of
\begin{equation}\label{PolV}
\hat{e_{n}}= \frac{k_n}{k_{n\perp}\sqrt{2}}[\hat{e}_z \times \hat{k_n}+i\hat{k_n} \times (\hat{e}_z \times \hat{k_n})]
\end{equation} 
Notice that 
The principal property that makes  the  Beltrami vectors especially useful in analyzing Nonlinear Alfv\'enic States is that they are the eigenstates 
of the curl operator 
\begin{equation}\label{Def BV}
\nabla \times {\bf Q}_n = k_n {\bf Q}_n
\end{equation}
with the eigenvalue $k_n$; the differential operator is converted into a number. Since $\hat{k_n}$ is perpendicular to both $\hat{e}_{n1}$ and $\hat{e}_{n2}$, the BV are, trivially, divergence free:
\begin{equation}\label{DivFree}
\nabla \cdot {\bf Q}_n= 0,
\end{equation}
and form an orthonormal basis,
\begin{equation}\label{Ortho}
{\bf Q}_n \cdot {\bf Q}_n = 0, \quad {\bf Q}_n \cdot {\bf Q}^*_n = 1
\end{equation}
 
The aforementioned completeness properties allow any divergence-free vector ${\bf G}$ to be expanded in a Beltrami basis
\begin{equation}\label{Beltexpand}
{\bf G} = \sum_n  G_n {\bf Q}_n
\end{equation}
where $G_n$,
\begin{equation}\label{Belttrans}
G_{n} = \int {\bf G} \cdot {\bf Q}^*_n d{\bf x} 
\end {equation}
are the Beltrami coefficients (transforms). 
Through the following steps, one demonstrates that the Beltrami transform is a mathematically consistent and invertible object (like the cousin Fourier)
\begin{equation}
\sum_{m} G_{m} \int {\bf Q}_{m} \cdot {\bf Q}_n^* d{\bf x}=
\sum_{m} G_{m} \hat{e}_{m} \cdot \hat { e}_n^* \int d{\bf x} e^{i ({\bf k}_m - {\bf k}_n)\cdot {\bf x}}  
= \sum_{m} G_{m} {\hat e}_{m} \cdot {\hat e}_n^* \delta_{m, n} = G_{n}
\end{equation}
Further, the Reality of ${\bf G}$ ( ${\bf G} = {\bf G}^*$) is insured by the relation,
\begin{equation}
 \sum_{n} G_n \hat {e_n} e^{i {\bf k}_n \cdot {\bf x}} = \sum_{n} G_n^* {\hat e_n}^* e^{-i {\bf k}_n \cdot {\bf x}} \implies { G}_n = -{G}_{-n}^*,
\end{equation}
which follows by changing the index in the second sum from n to -n, and using the properties: $\bf{k}_{-n}= -\bf{k}_n$ and ${\hat e}_{-n}^*=-{\hat e}_{n}$. Notice the difference ( negative sign) from the equivalent reality constraint on Fourier coefficients. 

It is important to state that the Beltrami expansion pertains only to divergence free fields and  limits the analysis to incompressible extended MHD. In fact there are further additional constraints that restrict the class of vector functions that allow a Beltrami expansion \cite{footnote}. 

\section {Beltrami-transformed hall magnetohydrodynamics}
The preceding review sets the background for exploring what may, perhaps, qualify as a new alternative formulation of  the theory of Nonlinear Alfvenic Motions (including turbulence). At the end of a somewhat straightforward derivation, we believe that some of the advantages of casting the theory in  a Beltrami  basis will become manifest; we will end up with a simpler  and more transparent set of final equations for the incompressible Alfvenic systems of considerable complexity (Extended MHD even with electron dynamics). In the main body of this paper, however, we will concentrate on Hall MHD (HMHD), governed by the normalized equations,
\begin{equation}\label{HMHD1}
\frac{\partial {\bf b}}{\partial t} -\frac{\partial}{\partial z} ( {\bf v} - ( \nabla \times {\bf b}) ) = \nabla \times [ {\bf v} \times {\bf b}  - (\nabla \times {\bf b} ) \times {\bf b} ] 
\end{equation}
\begin{equation}\label{HMHD2}
\frac{\partial \nabla \times {\bf v}}{\partial t} -\frac{\partial}{\partial z} (  \nabla \times {\bf b} ) = \nabla \times [ {\bf v} \times (\nabla \times {\bf v})  + (\nabla \times {\bf b} ) \times {\bf b} ] 
\end{equation}
describing a charged fluid (no electron inertia) embedded in a uniform magnetic field  along the z direction ( ${\bf B}_0= B_0 \hat{e_{z}}$). The variables 
{\bf b} and {\bf v} are the varying magnetic and velocity fields, the former normalized to $B_0$ and the latter to the Alfven speed $V_A=B_0/\sqrt{4\pi \ Mn_0}$ where $M$ is the ion mass and $n_0$ is the uniform density. The space and time scales are measured, respectively, in terms of the inverse of cyclotron frequency $(\Omega = qB_0/Mc)$, and the ion skin depth $\lambda= V_A/\Omega=c/\omega_{pi}$.

The left hand side of Eqs.(\ref{HMHD1})-(\ref{HMHD2}) contain only the linear terms while the nonlinear couplings make up the right hand side. 
The Beltrami transform  of  the HMHD Eqs.(\ref{HMHD1})-(\ref{HMHD2}) can be worked out following the conventional Fourier expansion route. The  transformed equations take the form (detailed derivation is given in Appendix.1), 
\begin{equation}\label{BHMHD1}
\frac{\partial b_m}{\partial t} - i k_{mz} ( v_m - k_m b_m )  = i \sum_{n} \left [v_n b_{m-n} - v_{m-n} b_n  + (k_{m-n} - k_n ) b_n b_{m-n} \right ] I_{nm} 
\end{equation}
\begin{equation}
k_m \left [ \frac{\partial v_m}{\partial t} - i k_{mz} b_m \right ] = i \sum_{n} (v_n v_{m-n} - b_n b_{m-n} ) ( k_{m-n} - k_n  ) I_{nm}
\label{BHMHD2}
\end{equation}
where : 1) the vector fields  (${\bf v}, {\bf b}$) have been replaced  by the scalar Beltrami amplitudes ($v_n, b_n$) and, 2) all the ${\bf k}$ dependent vectorial complications of the nonlinear terms (NL) are reduced to a rather compact and relatively simple scalar interaction kernel
\begin{align}
I_{nm} = ({\bf k}_n  \cdot \hat{e}_{m-n} )(\hat{e}_n \cdot \hat{e}_m^*).
\label{inm}
\end{align}
It is, perhaps, helpful to point out that  $(v_n b_{m-n} - v_{m-n} b_n)$,  representing  ${\bf v}\times {\bf b}$, one of the basic MHD nonlinearity, is the only term that is not multiplied by some $k$; all other nonlinearities have an extra curl in them. In principle, it should be the dominant nonlinearity in the $k<<1$ regime. However we will find that in most cases of interest, it is not so.

\subsection {The Interaction Kernel}

To begin a general delineation and discussion of the NL terms, we must explicitly evaluate $I_{nm}$.  In homogeneous media, one may be tempted to split the wave vector ${\bf k}_n = k_{nz} \hat{e}_z + k_{n\perp} \hat{e}_\perp=k_n[\hat{e}_z \sin{\theta_n}+ \cos{\theta_n}\hat{e}_\perp] , \quad k_n=\sqrt{k_{nz}^2+k_{n\perp}^2}$
 where $\hat{e}_z$ (direction of the ambient magnetic field) is the only characteristic system vector.  However, to properly describe a 3D nonlinear state, one must insist on a more general ( Fig.1) representation:
 \begin{equation}\label{kgeneral}
{ \bf k}=k_z \hat e_z +k_x \hat e_x+k_y \hat e_y\equiv k_z \hat e_z +{\bf k_{\perp}}\equiv \hat k k, \quad k=|{\bf k}|
\end{equation}
 in terms of which, $I_{nm}$ is evaluated in Appendix.2. Schematically, 
  \begin{equation} \label{polproductT}
 \hat{e}_n \cdot \hat{e}_m^*= \frac{X+iY}{2}, \quad {\bf k}_n  \cdot \hat{e}_{m-n}=\frac{G+iH}{\sqrt2} 
 \end{equation}
leading to
 \begin{equation}\label {J}
 i I_{nm}= \frac{1}{2\sqrt2} [-(XH+YG)+i(XG -YH)]\equiv [-A_{nm} +iB_{nm}]= J_{nm} e^{-i\zeta_{nm}}
\end{equation}
alongwith
\begin{equation}\label {J}
 i I_{mn}=[-A_{nm} - iB_{nm}]=J_{nm} e^{i\zeta_{nm}}
\end{equation}
where $J^2_{nm}=A_{nm}^2+B_{nm}^2$ and $\zeta_{nm}=\arctan(B_{nm}/A_{nm})$, and $I_{mn}=I_{nm}^*$.
 
The defining functions (X, Y,  G and H ), in complete generality, are 
 \begin{equation} \label{polstep3T}
X=(1+\hat k_m \cdot \hat{k_n}) \frac{{\bf k_{n\perp}} \cdot {\bf k_{m\perp}}}{k_{m\perp} k_{n\perp}}+ [\hat e_z\cdot({\bf k_n} \times {\bf k_m})]^2, 
\end{equation}

\begin{equation} \label{polstep4T}
Y=(\hat{k_n}+\hat{k_m})\cdot\hat e_z [\hat e_z\cdot{\bf k_n} \times {\bf k_m})],
\end{equation}

\begin{equation}\label{hybridK6T}
 G= \frac{[\hat e_z\cdot{\bf k_n} \times {\bf k_m})]}{K_{\perp}}, \quad  H=\frac{[{\bf k_{m\perp}}k_{nz}-{\bf k_{n\perp}}k_{mz}]\cdot{\bf K}_{\perp}}{KK_{\perp}}
\end{equation}
Notice the appearance of the modulus ($K$) of the hybrid (difference)vector 
 \begin{equation}\label{hybridKT}
{\bf K}\equiv{\bf K}_{m-n}={\bf k}_m -{\bf k}_n, K=|{\bf K}|,  K^2=k_m^2+k_n^2-2 {\bf k}_m \cdot{\bf k}_n ,\quad
{\bf K}_\perp={\bf k}_{m\perp} -{\bf k}_{n\perp}
\end{equation}
in the denominators of G and H; it will have a profound effect in boosting up the effects of nonlinearities when ${\bf k_n}$ and ${\bf k_m}$ are near parallel and have nearly equal magnitudes.
 
The expressions for $I_{nm}$ (displayed in the preceding equations), when injected into (\ref{BHMHD1})-(\ref{BHMHD2}), will yield a fully described Nonlinear Alfvenic system ready for numerical simulations. Needless to say  $I_{nm}$ is the central element in the current formulation of the Nonlinear Alfv\' enic dynamics expressed in the Beltrami basis; it is an exact consequence of HMHD equations. In  fact,  a very similar  set of equations emerge even when the electron dynamics is included. 

\begin{figure}
\includegraphics[width=4in]{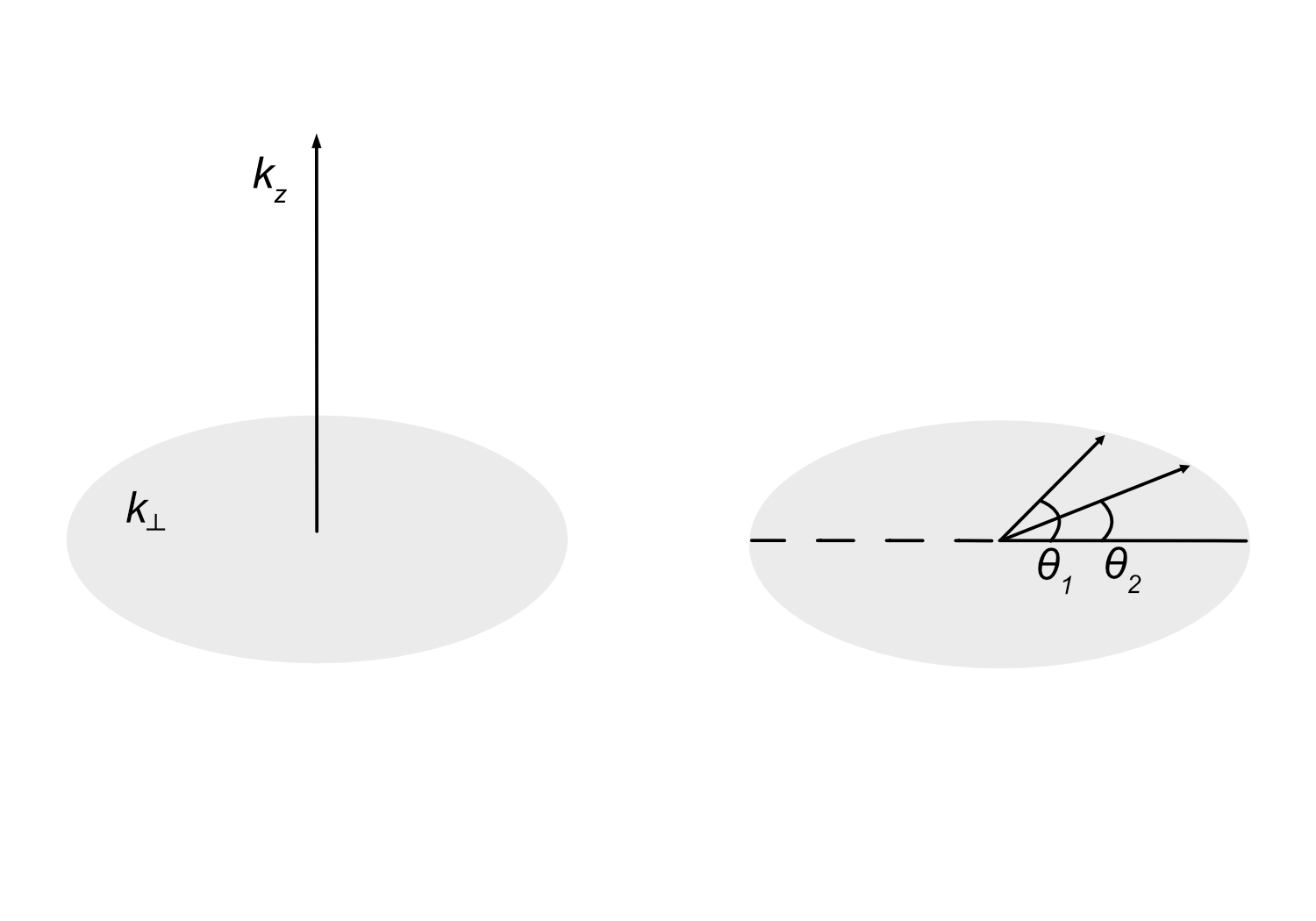}
\caption{The k-space}
\end{figure}

\subsection{General Characteristics of $I_{nm}$}

The main thrust of this paper, however, is to advance the theoretical-conceptual understanding of the Alfv\'enic system.  Specifying the ${\bf k}$ geometry will helps us in deriving results that are explicit and more transparent. Since the ambient magnetic field provides us with a unique  characteristic direction, we choose a cylindrical decomposition
\begin{equation}\label{kgeneralcyl}
{\bf k}=k_z \hat e_z +k_x \hat e_x+k_y \hat e_y= k_z \hat e_z + k_{\perp}(\hat e_x \cos\theta+\hat e_y \sin\theta)\equiv
k_z \hat e_z +{\bf k_{\perp}}
\end{equation}
where ${\bf k}$ is represented by the parallel ($k_z$) and perpendicular ($k_{\perp}$) projections, and the polar angle $\theta$. Invoking (\ref{kgeneralcyl}), explicit formulas for the constituents of $I_{nm}$ are derived in  Appendix 2A. We will refer to them as we proceed in our investigation of a model problem. However, before we do so, 
we must  draw the reader's attention towards several essential features of the nonlinear interactions:
\item the Interaction Kernel $I_{nm}$vanishes when the interacting vectors are parallel: ${\bf k_m}=\mu {\bf k_n}$. The functions Y and G are obviously zero and so is H; this is true independent of their magnitudes. 
 \item In general, the interaction kernel is complex with a real (imaginary) part that is symmetric (antisymmetric) on index exchange: $ I_{n m}=I_{m n}^*$. 
\item We will later find that the factor H is a sensitive function of $k$ lending extra structure to the interaction; it contributes an additional zero when $k_{n\perp}=k_{m\perp}$. This will become explicit in our model example.
\item Because of the factor $K^2= |{\bf k}_m - {\bf k}_n|^2$ in the denominator (especially of H), $I_{nm}$ will be highly boosted for wave vectors that are near parallel and near equal. This conforms to the expectation that two near-neighbor modes in k space will strongly influence one another. However, the $I_{nm}$ is sufficiently complicated (due to the aforementioned zeros) that we must await further calculations to learn the detailed structure. 

The preceding three sections, along with Appendices 1 and 2 constitute  Part I of this paper- the exact formulation of the fully nonlinear Hall MHD. This system is ready for a program of numerical simulations. The formalism can readily be extended to include the electron dynamics. We are now ready, and well- equipped to embark on Part II of the paper.

Part II

\section {Analyzing Nonlinear HMHD - Successive Approximations} 

The Beltrami transformed set [Eqs.(\ref{BHMHD1})-(\ref{BHMHD2})] is the basic set on which we will build a new approach for elucidating  the nonlinear (coherent or turbulent) Alfv\'enic states. It is simple enough to serve as a basis for numerical simulation. Whether this system is actually more efficient than the canonical Fourier system will be examined in a forthcoming work. As mentioned earlier, this paper is devoted, mostly, to analytic manipulation of the system to extract general results that are not only interesting in their own rights, but will also provide guidance and check to simulations. I will also demonstrate how this analysis could be relevant to understanding/ interpreting  phenomena observed in basic laboratory experiments on Alfv\'en waves.

In the first section of PartII, we will now develop an approximation for the nonlinear HMHD by exploiting an essential feature of the linear HMHD, namely, the relationship between the field amplitudes. To begin with, we  manipulate  (\ref{BHMHD1})-(\ref{BHMHD2}) by taking the time derivative of  (\ref{BHMHD1}), and then using (\ref{BHMHD2}) to eliminate the velocity field in the linear terms; we find
 \begin{equation}\label{CombHMHD}
(\frac{\partial^2}{\partial t^2}+ik_{mz}k_m+k_{mz}^2) b_m\equiv D_m b_m = i \frac {\partial N_1}{\partial t}+ i\frac{k_{mz}}{k_m}N_2
 \end{equation}
where the nonlinearities have been, schematically, denoted by $N_1$ and $N_2$.

\subsection{ Linear theory - a short review}
The linear theory as well as the so-called linear-nonlinear waves have been extensively investigated earlier; only a brief summary is provided here. The reader may consult earlier publications in this topic, e.g., \cite{W1,SV05,MM09,XZ15,ALM16,AY16}.

When the nonlinear terms are neglected, the system splits into two independent circularly polarized linear waves; each of them oscillates with a characteristic frequency $\omega_m$, determined by the dispersion relation,
\begin{equation}\label{Disp1}
D_m(\omega_m, k_m)\equiv {\omega_m}^2+k_{mz} k_m \omega_m - {k_{mz}}^2=0
\end{equation}
yielding
\begin{equation}\label{Disp2}
\frac{\omega^{\pm}_m}{k_{mz}}=\frac{k_m}{2}\pm \sqrt{1+\frac{k^2_m}{4}}\equiv -\alpha^{\pm}_m.
\end{equation}
The linear theory, in addition, fixes the ratio of the magnetic and kinetic amplitudes
\begin{equation}\label{Ampratio}
b_{m}=\alpha_m \\ v_{m};
\end{equation}
$\alpha_m=\alpha_m (k_m)$ is a function only of the modulus of the wave vector. The linear properties of the two branches $(\pm)$  will be rather important for our later nonlinear analysis. The dispersion relation has two branches: the ${\omega^{+}_m}$ corresponds to the Alfv\'en-Whistler  (AW) mode while the ${\omega^{-}_m}$ may be called the Alfv\'en- Cyclotron (AC). In the MHD limit $(k\rightarrow 0)$, these two modes reduce to $\omega^{\pm}_m/k_{m z}\rightarrow \pm1$, but the magnitude of their frequencies become very different when $k \geq 1$; the AW frequencies increases with k while that of AC approach a value that is $k_z/k$ times the cyclotron frequency. Naturally the nonlinear interactions of these two branches will be quite different. To keep track of  the two branches each with two $k$ limits (for analytical simplicity), I will use the following notation: defining $AW\equiv +, AC\equiv -$, a pertinent physical attribute X will have four manifestations: $X^{+}_{<},  X^{+}_{>}, X^{-}_{<}$, and $X^{-}_{>}$ where $<(>)$ denote the $k<<1(k\geq1)$ part of the spectrum. For example, the very important function ${\alpha}$, will be, explicitly,
\begin{equation}\label{ALPHA}
\alpha^{+}_{<}\approx -(1+\frac{k}{2}), \quad \alpha^{+}_{>}\approx -k , \quad \alpha^{-}_{<}\approx 1-\frac{k}{2},  \quad \alpha^{-}_{>}\approx \frac{1}{k}
\end{equation} 
We will use this complicated notation only when it is needed to avoid confusion.

\subsection{ Approximate HMHD- Modulated nonlinear Alfv\'en Waves}

This recapitulation of the linear theory gets us ready to rewrite the master equation (\ref{CombHMHD}) as, 
 \begin{equation}\label{CombHMHDL}
D_m b_m =  \sum_{n}  iI_{nm}  [g_{nm}\frac{\partial (b_n b_{m-n})}{\partial t} +  i k_{mz} f_{nm}(b_n b_{m-n})]
\end{equation}
with
\begin{equation}\label{gandf}
g_{nm}= \frac{1}{\alpha_n}  - \frac{1}{\alpha_{m-n}}  + k_{m-n} - k_n,  \quad    f_{nm}= (\frac{1}{\alpha_n\alpha_{m-n}}-1)  \frac{(k_{m-n} - k_n)}{k_m},
\end{equation}
after eliminating the velocity field via $v_m=b_m/\alpha_m$, where $\alpha_m=\alpha_m(k_m)\equiv -\omega_m/ k_{zm}$ is just an alternative expression of the dispersion relation, it is a real function of the wave momentum.  It is, perhaps, pertinent to emphasize that (\ref {CombHMHDL}) has all the content of the original system but for the  simplification introduced by using  $v_m=b_m/\alpha_m$ (only) in the nonlinear terms; Equation (\ref {CombHMHDL}), therefore, could serve as a good but somewhat approximate/easier model for numerical simulations of HMHD.

Deriving and studying an analytically tractable system, that may be aptly called Modulated nonlinear Alfv\'en Waves, is the main focus of PartII.  Such a model must, necessarily, involve a small/finite set of interacting modes. For the time being, however, we will proceed generally. The modulated nonlinear Alfv\'en Waves form that class of solutions where the nonlinearities affect the linear waves on time scales that are longer than the inverse linear frequencies.  Without nonlinear coupling, the time behavior of the independent  linear HMHD waves would have been of the type
\begin{equation}\label{Lin}
 b_{m}= \mathrm{const} \cdot e^{-i\omega_m t},  
\end{equation} 
but with nonlinear interactions, we must seek a solution of the type
 \begin{equation}\label{LinT}
 b_{m}= \psi_m (t) e^{-i\omega_m t}
 \end{equation}
where $\psi_m (t)$ represents the time history of the modulation induced by nonlinear coupling, and will have the information on how the modes exchange energy. It is quite reasonable to assume that the time rate of change 
\begin{equation}\label{SlowMod}
 \frac{1}{\psi_m (t)} {\frac{d\psi_m (t)}{dt}} \ll \omega_m;
 \end{equation}
is smaller than the wave frequency; the time dependence of  the envelope function $\psi_m (t)$  will represent the modulation of a rapidly oscillating field. 

Substituting (\ref{LinT}) into (\ref{CombHMHDL}), and using the linear dispersion relation (to simplify the the linear part), we arrive at the (approximate) modulation equation
\begin{equation}\label{EqnEvolution}
\frac {d\psi_m}{dt}= -\frac {1}{\pm 2\sqrt{1+k_{m}^2/4} }\sum_{n}  iI_{nm}  [\alpha_m g_{nm} +  f_{nm}] \psi_{n} \psi_{m-n}:
\end{equation}
provided the resonance condition 
\begin{equation}\label{ResonantCondition}
\omega_m=\omega_n +\omega_{m-n},
\end{equation}
which eliminates the fast variation(on Alfv/'en times), is satisfied. Each modulated Beltrami amplitude $\psi_m$ is being driven by three wave processes; all combinations of the class $\psi_n \psi_{m-n}$ generate $\psi_m$. Immensely simplified, the system is still too  cumbersome to allow analytic inroads. 

It is essential that we understand the implications and constraints of the resonance condition. When supplemented by the exact dispersion relation, the resonance condition (\ref{ResonantCondition}) reads
\begin{equation}\label {ResonantConditionFull}
k_{mz} ( \frac{k_m}{2}\pm \sqrt{1+\frac{k^2_m}{4}})=k_{nz} ( \frac{k_n}{2}\pm \sqrt{1+\frac{k^2_n}{4}})+(k_{mz}-k_{nz}) ( \frac{K}{2}\pm \sqrt{1+\frac{K^2}{4}})
\end{equation}
where $k_{{(m-n)}z}\equiv k_{mz}-k_{nz}$ and $K$ stands for $|{\bf k_m}-{\bf k_n}|$, the modulus of the difference /hybrid vector. the dispersion relation, thus, transforms the frequency matching constraint into one that dictates wave number matching, albeit in a nontrivial form.  

In its entirety, the relation (\ref{ResonantConditionFull}) is complicated as well as opaque. It begins to make sense, however, when we examine it in more familiar limits. Because of the richness of Alfvenic spectrum, there are several combinations that arise. 
\subsection {Low k MHD Regime} 
We will first examine the low k MHD limit in the following obvious classes of interactions:
\item All three interacting modes are on the Alfv\'en-Whistler branch (+++)
\begin{equation}\label {ResonantConditionFullWhi}
k_{mz} ( \frac{k_m}{2}+1)=k_{nz} ( \frac{k_n}{2}+ 1)+(k_{mz}-k_{nz}) ( \frac{K}{2}+1)\quad \Rightarrow \quad \frac {k_{mz}}{k_{nz}}
=\frac {k_n-K}{k_m-K}
\end{equation}
\item All three interacting modes are on the Alfv\'en-Cyclotron branch (---)
\begin{equation}\label {ResonantConditionCycl}
k_{mz} ( \frac{k_m}{2}-1)=k_{nz} ( \frac{k_n}{2}- 1)+(k_{mz}-k_{nz}) ( \frac{K}{2}-1)\quad \Rightarrow \quad \frac {k_{mz}}{k_{nz}}
=\frac {k_n-K}{k_m-K}
\end{equation}
The wave number constraints for the two are exactly the same. It happens because the leading order terms (linear in k), trivially, cancel and the constraint comes from the quadratic terms which are exactly the same.

\item Two modes are on one branch while the third is on the other. It turns out  that in the model problem, defined through (\ref{EqnEvolution}), all such combinations $[(++-) (+-+)(--+) (-+-)]$, to the leading order, yield the same essential result: One of $k_z$s, namely, $k_{mz}$, $k_{nz}$, or $K_z=k_{mz}-k_{nz}$ must be zero implying (respectively) that either $\omega_m$, $\omega_n$ or $\omega_{m-n}$ must be zero. Since this will violate our assumption (\ref{SlowMod}), these intra-branch processes should not contribute to the evolution of $\psi_m$ via Eq. (\ref{EqnEvolution}). Strictly speaking it is possible (though hard) to satisfy the resonance condition if one includes terms to the next order in $k$; one must, then, require
\begin{equation}\label {ResonantConditionFullWhi}
k_{mz} ( \frac{k_m}{2}+1)=k_{nz} ( \frac{k_n}{2}+ 1)-(k_{mz}-k_{nz}) ( \frac{K}{2}+1)=>\frac{k_{mz}}{k_{nz}}=\frac {k_n+2+K}{k_m+2+K}
\end{equation}
which give a small but finite value for 
\begin{equation}\label {ResonantConditionFullWhi}
\frac{k_{mz}- k_{nz}}{k_{nz}}\approx \frac{k_n-k_m}{2}<<1  
\end{equation}
In fact one could explore situations where, for instance, the $[k_1, k_2 > 1]$ while the $K<1$. Thus the cross mode couplings could be analyzed within this framework. But to limit the paper to a finite size and preserve physical transparency, we will not explore them here.   

It must be emphasized that the contribution of even $k_z=0$ (zero frequency modes) to the general class of interactions in the original system (\ref{BHMHD1})-(\ref{BHMHD2}) [or even (\ref{CombHMHDL}) for that matter] is not forbidden. It is just not allowed for the restricted class comprising the Modulated Nonlinear Alfv\'en Waves because of the imposed separation in the time scales.

\section{A Truncated Analytical Model for Modulated Waves- Basics}
We will now explore a subset of the system (\ref {EqnEvolution})-(\ref {ResonantCondition}) to gain  deeper theoretical understanding of some of the processes that will control the  formation of an eventual Nonlinear Alfv\'enic State.  Let us suppose that we "prepare" a system in which only three modes $(\psi_{m_1}, \psi_ {m_2}, \psi_{m_2-m_1}=\psi_h)$ associated with ${\bf k}_{m_1}$, ${\bf k}_{m_2}$, ${\bf k}_{m_2}-{\bf k}_{m_1}={\bf K}$, along with their conjugates  ($\psi_{-m_1}= -\psi^*_{m_1},  ---)$, are excited. In order to avoid clutter in the notation, let us  introduce $1, 2$ and $h\rightarrow 2-1$ as a shorthand notation for $m_1, m_2$, and $m_2-m_1$; thus it is important to appreciate  that $2-1\neq 1$. The envelope dynamics of this system is controlled by Eqs. (\ref{EqnEvolution1}) -
(\ref{EqnEvolutionh}) derived in Appendix.3. 
\begin{equation}\label{EqnEvolution1T}
\frac {d\psi_1}{dt}= \frac {J_{21}} {2}[\alpha_1 g_{21} +  f_{21}] \psi_{2} \Psi_{h}\equiv \frac{\mu_1J_{21}}{2} \equiv u_1\psi_{2} \psi_{h}
\end{equation}
\begin{equation}\label{EqnEvolution2T}
\frac {d\psi_2}{dt}= \frac {J_{21}}{2}[\alpha_2 g_{12} +  f_{12}] \psi_{1} \psi_{h}\equiv \frac{\mu_2 J_{21}}{2}\equiv u_2\psi_{1} \psi_{h}
\end{equation}
\begin{equation}\label{EqnEvolutionhT}
\frac {d\psi_h}{dt}= \frac { J_{2h}-  J_{-1h}}{2}[\alpha_h g_h +  f_h] \psi_{1} \psi_2\equiv \frac{\mu_h (J_{2h}-J_{- h})}{2}\equiv u_h\psi_{2} \psi_{1}
 \end{equation}
where the $[u_1= u_1(k), u_2= u_2(k), u_3= u_3(k)]$ are rather  complicated functions of the wave vectors, through the $J$s 
and through the $\mu$s. In the main text, we will work out in detail only the low $k$ MHD-like limit of the plus mode. In this limit, the $\mu$s, given by,
\begin{equation}\label{Mu1}
\mu_1=\alpha_1 g_{21} +  f_{21}=-\frac{K-k_2}{2k_1}[k_1+k_2+K],
\end{equation}
\begin{equation}\label{Mu2}
\mu_2=\alpha_1 g_{12} +  f_{12}=-\frac{K-k_1}{2k_2}[k_1+k_2+K],
\end{equation}
\begin{equation}\label{Muh}
\mu_h=\alpha_h g_{h} +  f_{h}=-\frac{k_1-k_2}{2K}[k_1+k_2+K],
\end{equation}
are totally symmetric. 

The analysis of this this system begins by noting that since all $u$ are independent of time, Eqs.(\ref{EqnEvolution1T})-(\ref{EqnEvolutionhT}) are always reducible to appropriate elliptic integral, or expressed in terms of Jacobian Elliptic functions. This is explicitly shown in Appendix 4 (which has additional details). However, in the main text, we will limit ourselves to working out limiting cases that are readily integrated (in terms of elementary functions) but that capture the essentials of the triplet dynamics.

The solutions of (\ref{EqnEvolution1T})-(\ref{EqnEvolutionhT}) can be very diverse because their  nature will be set by the relative signs of the coupling coefficients $u$. And naturally the sign would change as we move about in the ${ k_1, k_2}$ space. it turns out that for most regions in the said space, the J  functions have the same form and sign; the relative  $\mu$ signs, then, dictate the relative $u$ signs. We will further assume $J$ s to be positive (the entire calculations will be the same if they were all negative). That places a helpful limit on the number of independent cases that we must study. A further limit is imposed by the symmetry of the system. 

The first step in calculation is to, formally, set $u_i=\sigma(\mu_i) |u_i|$ ($\sigma(\mu)$ is the sign of $\mu$), define a new normalized time
\begin{equation}\label{NormT}
\tau=t\sqrt{|u_1| |u_2| |u_h|},
\end {equation}
and rewrite the evolution equations as  ( $ \psi_i \rightarrow \Psi_i \sqrt{u_i}$)
\begin{equation}\label{parameterfree}
 \frac {d\Psi_1}{d\tau}= \sigma(\mu_1) \Psi_{2} \Psi_{h}, \quad  \frac {d\Psi_2}{d\tau}= \sigma(\mu_2)\Psi_{h} \Psi_{1},\quad \frac {d\Psi_h}{d\tau}= \sigma(\mu_3) \Psi_{1} \Psi_{2},
\end {equation}
a system without any free parameters. Going from (\ref{EqnEvolution1T}) - (\ref{EqnEvolutionhT}) to (\ref{parameterfree}), though algebraically trivial, conveys a fundamental message;  independent of the time history of the evolution, the only effective time intrinsic to the triple dynamics is $\tau$, fully controlled by the product  $|u_1| |u_2| |u_h|$ of the coupling strengths. The stronger the coupling, the faster the nonlinear processes.  

To this I must add a word of caution. For a nonlinear initial value problem, the initial conditions may (and indeed do) modify the time scales. We will soon see, how? 

Without any loss of generality, we will assume $k_2>k_1$ and set
\begin{equation}\label{NormT}
\sigma(\mu_3)= +1
\end {equation}
The evolution of the modulated waves in a given $[k_1, k_2, K]$ region will, then, depend on the corresponding values of $\sigma (\mu_{1, 2})$. Three distinct classes of dynamics emerge: 1) Strong  directional (in k) energy transfers amongst the participating modes, 2) periodic energy exchanges, and  3) No Direct energy exchange-essentially independent mode trajectory. 

\subsection{Strong Energy Exchange Regimes - $k_2>k_1$}
We will now show that strong energy exchanges take place when:  

1) $K>k_2>k_1$ - the hybrid $K$ is the largest wave number, 

2) $k_1<K<k_2$ - $k_2$ is the largest wave number

For the first [second] ordering, $\sigma (\mu_1)=-1=\sigma(\mu_2)$ [ $\sigma(\mu_1)=1, \sigma(\mu_2)=-1$], leading, respectively, to 
\begin{equation}\label{pf--+}
 \frac {d\Psi_1}{d\tau}= - \Psi_{2} \Psi_{h}, \quad  \frac {d\Psi_2}{d\tau}= -\Psi_{h} \Psi_{1},\quad \frac {d\Psi_h}{d\tau}=  \Psi_{1} \Psi_{2}
\end {equation}
and
\begin{equation}\label{pf+-+}
 \frac {d\Psi_1}{d\tau}= \Psi_{2} \Psi_{h}, \quad  \frac {d\Psi_2}{d\tau}= -\Psi_{h} \Psi_{1}, \quad \frac {d\Psi_h}{d\tau}=  \Psi_{1} \Psi_{2}
\end {equation}
We will detail all steps for solving (\ref{pf--+}); the solution of (\ref{pf+-+}) is readily constructed by analogy. Equations  (\ref{pf--+})
allows two conserved (energy) combinations
\begin{equation}\label{Energy1-2T}
\Psi_{1}^2+ \Psi_{h}^2=E, \quad  \Psi_{2}^2+ \Psi_{h}^2= aE \quad \Rightarrow \Psi_{1}= \sqrt{E - \Psi_{h}^2}, \quad  \Psi_{2}= \sqrt{aE - \Psi_{h}^2}  
 \end{equation}
where $E$ and $a$ are two positive constants. From (\ref{pf--+}) and (\ref{Energy1-2T}), one derives 
 \begin{equation}\label{FinAmplitude}
\frac{d\hat{\Psi_h}}{dT} =  \sqrt{1- \hat{\Psi_{h}}^2} \sqrt{a - \hat{\Psi_{h}}^2}
 \end{equation}
where ${\hat{\Psi_h}}= {\Psi_h}/\sqrt{E}$ is the normalized amplitude and 
\begin{equation}\label{effectivetimeE}
T={E^{\frac{1}{2}}} \tau=  (E|u_1| |u_2| |u_h|)^{\frac{1}{2}} t 
 \end{equation}
is the new effective time. This nonlinear rescaling of time is what was alluded to earlier. Since E is an effective initial energy, the time scales for nonlinear processes become shorter as the initial energy of the interacting modes is increased. Referring the reader to Appendix 4 for the exact Elliptic function  formulation, we will discuss here some special solutions of (\ref{FinAmplitude}). Notice that all acceptable solutions are bounded: $\hat{\Psi_{h}}^2< [1, a]$.

\item  Directional Energy Transfer:  the strongest energy transfer interactions take place when the constant $a=1$, i.e,  when the modes $\Psi_{1}$ and $\Psi_{2}$ begin with equal energies. The initial value  problem is, then,  solved by $(T>0)$, 
\begin{equation}\label {SOLF}
\hat{\Psi_h}=  \tanh (T\pm\chi_{h}), \quad \quad \hat{ \Psi_1}=\hat{\Psi_2}=  \mathrm{sech}(T\pm\chi_{h}),
\end{equation}
with initial values
\begin{equation}\label {SOLIn}
 \quad \hat{\Psi_h}(0)= \pm \tanh(\chi_{h}),   \quad \quad   \hat{\Psi_2}(0)= \mathrm{sech}(\chi_{h})
\end{equation}
where the upper(lower) sign pertains when modes h and [1, 2], initially, have the same (opposite )phase. In Figs(1-2) we have plotted the amplitudes [ $\hat{\Psi_h}, \hat{\Psi_2}$] and the energies [$\hat{\Psi_h^{2}}, \hat{\Psi_2^{2}}$] for the same and opposite phases taking the initial energies to be equal  
\begin{equation}\label {SOL}
\hat {\Psi}^{2}_{h}(0)= \hat{\Psi}^{2}_{2}(0) \Longrightarrow sech^{2}(\chi_h)= \tanh^{2}(\chi_h)=\frac{1}{2}
\end{equation}
The time trajectories of the interacting modes is displayed in Fig.2 (in phase initially)and Fig.3 (opposite phase initially).
\begin{figure}
\includegraphics[width=4in]{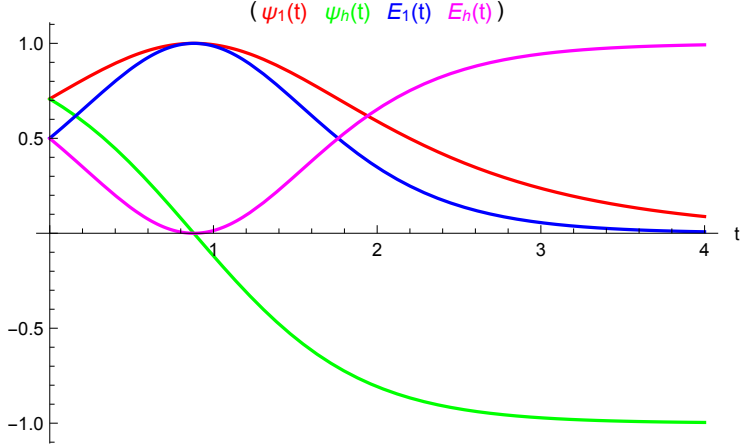}
\caption{The time evolution of the envelopes of  $\psi_{1}(t)=\psi_{2}(t), \psi_{h}(t)$  and $E_{1, h}=\psi^2_{1.h}$ starting from $E_1=E_h$ and $\psi_1$ and $\psi_2$ begin in phase. The time is measured in units of $t_{depletion}$.}
\end{figure}
\begin{figure}
\includegraphics[width=4in]{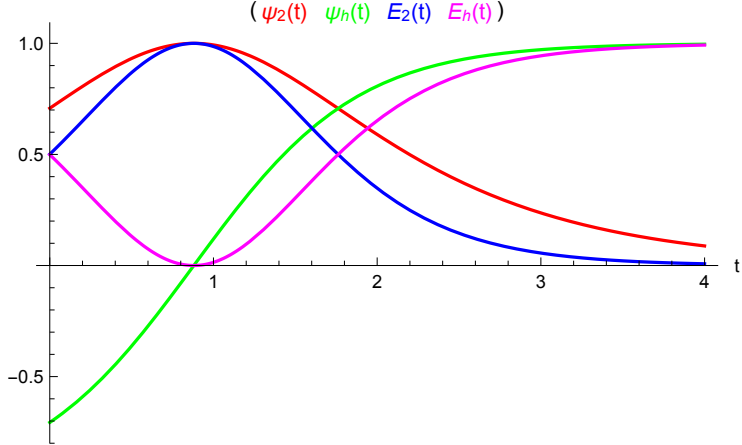}
\caption{The time evolution of the envelopes of  $\psi_{2}(t)=\psi_{1}(t), \psi_{h}(t)$  and $E_{2, h}=\psi^2_{2.h}$ starting from $E_1=E_h$ and $\psi_1$ and $\psi_2$ with opposite phase at t=0. The time is measured in units of $t_{depletion}$.}
\end{figure}

When the two modes are excited in phase (with equal energies), the set [$\hat\Psi_{2}(t), E_2= {\hat\Psi}^{2}_{2}(t)$] decrease monotonically reaching their asymptotic value 0, as [$ \hat\Psi_{h}(t), E_h= {\hat\Psi}^{2}_{h}(t) ]$ build up to [minus, plus] unity (Fig. 2). Thus, there is a continuous transfer of energy from the lower frequency [$\omega_2, k_2$] to the higher frequency mode $(\omega_h, K)$ till the former is fully depleted. The overall transfer takes place in units of, what could be, rightfully, labelled as the depletion time
\begin{equation}\label{effectivetimeE}
t_{depletion} =  (E|u_1| |u_2| |u_h|)^{-\frac{1}{2}},   
 \end{equation}
 which, in fact, defines the time constant of the (three wave) nonlinear processes on which the Alfv\'enic states evolve. Deriving revealing and succinct expression for the characteristic times like $t_{depletion}$ is, in fact, the central quest of this analytic investigation. Notice that  it is determined, not only by the interaction strengths (fully known as functions of $k$s), but also, by the initial energy in the wave system. 

As we can see in  Fig.(3), there is, more structure to the evolution of the amplitudes as well as energies. There is some initial period in which energy is transferred from  mode h to  mode 2] , but the final state is exactly the same as in Fig.2. It is easy to show that, independent of the initial distribution of the conserved energy $E=E_h+E_2$, the asymptotic state always settles to $E_h=>E$, $E_2=>0$

For the $k_1<K<k_2$ hierarchy, by following exactly the same procedure, one finds 
\begin{equation}\label {SOLF}
\hat{\Psi_2}=  -tanh (T\pm\chi_{h}), \quad \quad   \hat{ \Psi_1}=  \hat{ \Psi_1}=\hat{\Psi_2}=  sech(T\pm\chi_{h}),
\end{equation}
leading to the asymptotic transfer of all energy to mode [2] which, for this hierarchy, has the highest k: $E_t=>E$, $E_h=E_1=>0$. The qualitative features of the energy exchange processes in the two k regimes [ $K>k_2>k_1$,  2) $k_1<K<k_2$ ] may be summarized as:  

1) For all variations in the initial conditions (relative phases and magnitude of the initial amplitudes), it is the highest ($k$) mode that eventually ends up getting all the initial energy. This is a robust conclusion that holds no matter how the total energy is distributed in the beginning,

2) The basic characteristic time scale for energy exchange (substantial transfer of energy into the higher k mode)
 \begin{equation}\label{effectivetimeE1}
\frac{1}{t_{depletion}}=  \sqrt{(E|u_1| |u_2| |u_h|)} 
\end{equation}
is the fundamental global signature of the nonlinear modulation processes. 

3) The depletion time scales inversely with the initial energy ($\sim \sqrt E$), and  has complicated  $k$ dependence through the nonlinear interaction strengths $(\sim\sqrt{|u_1| |u_2| |u_h|})$ that drive each of  the triplets.  It is natural to expect that the stronger the interaction, and higher the initial energy, the shorter will be the time needed for energy exchange

\item Periodic Energy Exchange;
The full directional transfer of energy rakes place only in a very limited range of $a\approx1$. For most other values of a, the elliptic integrals reduce to periodic functions and the energy exchanges will be periodic. In fact a=1 is somewhat of a singular limit where the period becomes infinite and energy exchange is well represented by the monotonically increasing (decreasing) tanh (sech) function. Let us now work out the extreme sinusoidal limits of (\ref {FinAmplitude}) (redisplayed here for convenience)
\begin{equation}\label{FinAmplitude1}
\frac{d\hat{\Psi_h}}{dT} =  \sqrt{1- \hat{\Psi_{h}}^2} \sqrt{a - \hat{\Psi_{h}}^2}
 \end{equation}
The magnitude of all solutions of (\ref {FinAmplitude1}) are bounded below the lower of $[1, a]$. Two obvious limiting regimes emerge:

1) $a\ll1$ implying $\hat {\Psi_{h}}^2 \ll 1$, then (\ref{FinAmplitude1}) may be approximated as $(\hat{\Psi_h}(0)= 0)$,
\begin{equation}\label{FinAmplitude2}
\frac{d\hat{\Psi_h}}{dT} \approx \sqrt{a - \hat{\Psi_{h}}^2} \Rightarrow \hat{\Psi_h}= \sqrt{a}\sin(T), , \quad \hat{\Psi_2}=\pm \sqrt{a}\cos( T), \quad \hat{\Psi_1}\approx 1- a \sin^{2}(T)
 \end{equation}
A fraction $\approx a<<1$ of the initial energy is periodically exchanged between $[h]$ and $[2]$ while $[1]$ stays close to unity. The effective oscillation time is the same as $t_{depletion}$. 

2) In the limit, $a\gg1$  and $\hat\Psi_{h}^2<1$, (\ref {FinAmplitude1}) reduces to
\begin{equation}\label{FinAmplitude3}
\frac{d\hat{\Psi_h}}{dT} \approx \sqrt {a}\sqrt{1 - \hat{\Psi_{h}}^2} \Rightarrow \hat{\Psi_h}= \sin(\sqrt {a}T) , \quad \hat{\Psi_1}=\pm \cos( \sqrt{a}T), \quad \hat{\Psi_2}\approx a -  \sin^{2}(\sqrt{a}T)
 \end{equation}
Again a small fraction $(\sim 1<< a)$ of initial energy oscillates between mode$h$ and mode $1$ while mode $2$ remains close to $a\gg1$. For this latter energy regime, the effective oscillation rate
\begin{equation}\label{FinAmplitude3}
\frac{1}{t_{osc}}= \frac{1}{\sqrt{a}\quad t_{depletion}}
 \end{equation}
is larger than the depletion time relevant to processes that transfer energy to higher $k$ modes. It is larger by a factor of $\sqrt{a}$; this is because the new initial ``energy'' could be interpreted as being $a E$. The rescaling is consistent with (\ref{Energy1-2T}), and all timescales herein scale with the square root of the energy in question (in this case $a E$). Both these extreme limits are illustrated in Fig. 4; the oscillation period gets shorter for large a.

When $a$ is in vicinity of unity, the entire richness of  Jacobian Elliptic functions is on display (Fig.5). Concentrating on $\Psi_{h}$, one sees a $tanh$ like build up, then a flat top phase (near 1) before the functions begin its descent  into an oscillation. As $a$ gets closer and closer to unity, the flat top region stays put longer and longer approaching a pure $tanh$ solution. It will eventually turn around, but the turning time may be too large to be of interest.    

\item It is interesting to compare and contrast the two modalities of energy exchange, described above. Although the directional energy transfer happens only for a limited range of initial conditions (when the energies of the two of the three participating modes are initially very close), it is efficient and complete in that the entire energy of the lower $k$ modes is fully transferred to the highest k. In contrast, the periodic exchanges operate over a much larger range of initial conditions, the total oscillating energy is but a small fraction of the system energy; much of the initial energy just stays in the original mode. So I will contend that it is the former mechanism that may play the dominant role in determining the spectral distribution of the asymptotic state.

\begin{figure}
\includegraphics[width=4in]{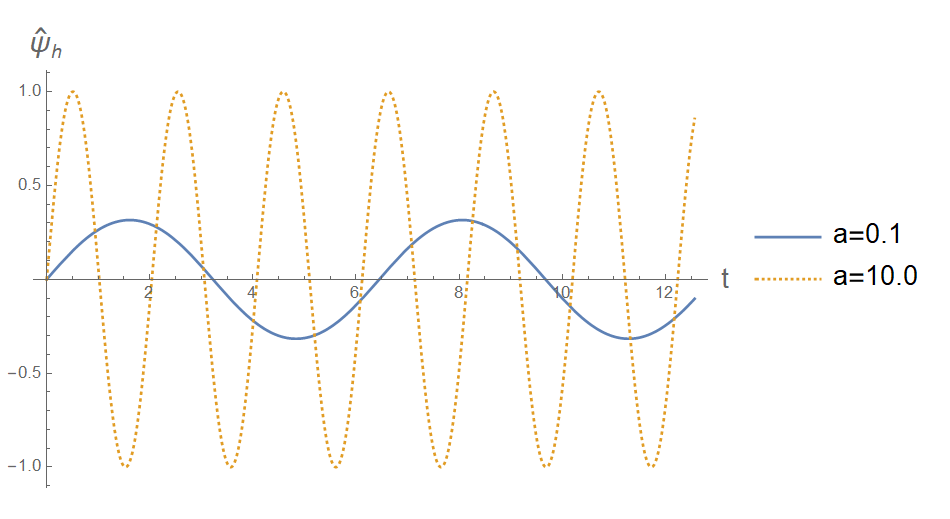}
\caption{The time evolution for the envelope $\hat{\psi}_h$ for $a<<1$ and $a>>1$. Both solutions represent an oscillatory sine wave. Notice that the periods are consistent with (66) and (67).}
\end{figure}

\begin{figure}
\includegraphics[width=4in]{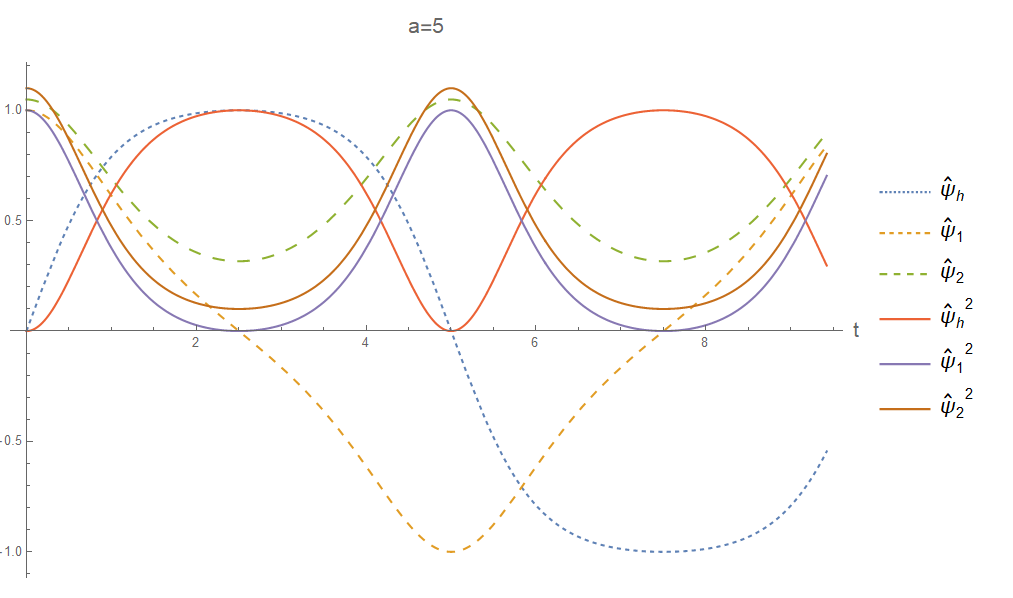}
\caption{The time evolution for the envelopes and their energies for the case when $a$ is very near but not exactly one. As $a$ approaches 1 the period of oscillation increases and the structure approaches a hyperbolic tangent structure characterized by a rapid change and a long saturation before periodically cycling.}
\end{figure}
\subsection {No Direct  Energy exchange}
The dynamics in the $k$ regime with ordering $K<k_1<k_2$ differs fundamentally from what we encountered in the orderings [$K>k_2>k_1$,  2) $k_1<K<k_2$] where energy exchange processes are dominant. Since  all $\sigma(\mu)$s are now positive, the triplet 
\begin{equation}\label{pf+++}
 \frac {d\Psi_1}{d\tau}=  \Psi_{2} \Psi_{h}, \quad  \frac {d\Psi_2}{d\tau}= \Psi_{h} \Psi_{1},\quad \frac {d\Psi_h}{d\tau}=  \Psi_{1} \Psi_{2},
\end {equation}
instead of relating mode energies through conserved sums, imposes the constraints  
\begin{equation}\label{MinimalExchangeEq}
\Psi_{1}^2= \Psi_{h}^2+
\mathcal {E},  \Psi_{2}^2= \Psi_{h}^2+p \mathcal {E} \quad \Rightarrow \Psi_{1}= \sqrt{\mathcal{E}+\Psi_{h}^2}, \quad  \Psi_{2}= \sqrt{p\mathcal{E} +\Psi_{h}^2}  
 \end{equation}
where $\mathcal{E}$ and $p$ could be positive or negative. Here it is the difference between the mode energies that remains constant; thus neither mode feeds off or decays onto another. Again leaving the formal elliptic function formulation to Appendix 4, let us explore some special cases. Let us, for instance, assume $p=1(\Psi_{1}=\Psi_{2})$; that leads to   
\begin{equation}\label {R1}
\frac{d\Psi_h}{d\tau}=  \mathcal{E}+\Psi_{h}^2
\end {equation}
For $\mathcal {E}>0$, (\ref{R1}) is solved by ($\Delta$ is a constant of integration)
\begin{equation}\label {R1}
 {\Psi_h} =  \sqrt{\mathcal{E}} \tan ({\sqrt\mathcal{E}}\tau+\Delta)
\end {equation}
which diverges when the argument reaches $\pi/2$- this solution is obviously not acceptable. For $\mathcal{E}< 0$, we must solve
\begin{equation}\label {R1}
\frac{d\Psi_h}{d\tau}=  - |\mathcal{E}|+\Psi_{h}^2
\end {equation}
with the restriction that $\Psi_{h}^2\geq |\mathcal{E}|$ to insure the reality of $\Psi_{1,2}$. The bounded well-behaved solution to the initial value problem is ($\Delta\geq 0$, $\tau$ positive)
\begin{equation}\label {R1}
 {\Psi_h} = - \frac{\sqrt{|\mathcal{E}|}}{ \tanh (\sqrt|{\mathcal{E}|}\tau+\Delta)}\Rightarrow {\Psi_h}^2= \frac{|\mathcal{E}|}{ \tanh^{2}(\sqrt|{\mathcal{E}|}\tau+\Delta)}, \quad {\Psi_{1, 2}}^2= {\Psi_h}^2 - |\mathcal{E}|.
 \end {equation}
The energy in the h mode decreases from its initial value $|\mathcal{E}| \tanh^{-2}(\Delta)$ to $|\mathcal{E}|$ as time advances while the energies of modes[1, 2] always trails by an amount $|\mathcal{E}|$ . Though coupled, no mode grows or damps at the cost of another.  Since the system is totally symmetric in [h,1,2], any permutation will have the same results.

 We have just demonstrated that the three-wave  resonant interactions (on time scales much slower than the wave frequencies) span an impressive range of nonlinear processes- ranging from ones with directional or periodic energy transfer to ones with no direct energy exchange. This Section covering the basics of the truncated analytical model (for modulated waves), though self contained, should be read in conjunction with Appendix. 4. 

\section {Analyzing Nonlinear  Truncated HMHD - a journey in the Wave number space}

The most involved part of the analysis is, obviously, the elucidation of the $k$ dependence of the exchange dynamics. Fortunately, the essence of the triplet dynamics is contained in a single physical parameter 
\begin{equation}\label{FinAmplitude3}
 \frac{1}{t_{depletion}}= (K+k_1+k_2)^{3/2}\left[\frac{|K-k_1| |K-k_2| |k_2-k_1|}{K k_1 k_2}\right]^{1/2} \left(|J^{2}_{12}(J_{2h}-J_{-1h})|\right)^{1/2} ,
\end{equation}
which, in principle, is fully known because the J coefficients have been evaluated in Sec.IIIA and Appendix.2 (Sec. IX).  For more explicit estimation, perhaps, the most straightforward (and quite easily doable) approach will be the numerical one where the  $J$ functions are calculated for a variety of  ${\bf k_1}, {\bf k_2}$, and the solutions plotted to illustrate discernible trends/ patterns.  This exercise will be left for a more detailed paper; for the time being we will make attempts to develop some insights by simple analytical examination of the interaction coefficients. 

Even before probing the $k$ structure of $J$, two important determinants of the depletion time (which, in fact, embodies  all relevant characteristic time scales) are evident:

1) $t_{depletion}$ decreases strongly with the sum of the three interacting wave numbers reminding us of the curl nonlinearities that dominate extended MHD

2) when any two of the wave numbers are close, the effective triplet interaction tends to diminish (vanishing at exact equality); consequently the energy exchange/oscillation time become larger and larger. The current model will, of course, cease to be accurate/valid when $t_{depletion}^{-1}$ approach the corresponding Alf\'ven frequencies.

\vspace {.3 cm}

Although I will comment on the mode interactions when the parallel and perpendicular wave numbers are 
comparable in magnitude, most of what is presented here, pertains to the spectral regime of particular interest to Alfv\'enic turbulence: it will be assumed that $k_z<<k_{\perp}\approx k$. In the cylindrical k representation of (\ref{kgeneralcyl}), this will limit the k space to a thin disk with its radius $(k_{\perp})$ much greater than the height ($k_z$). The setup is akin to the standard ``thin disk'' models used in astrophysics, e.g., when modeling the Milky Way.


Our discussion on the implications of the resonance condition $(\omega _2=\omega_1+\omega_h)$, as encompassed in (\ref{ResonantCondition}), (\ref {ResonantConditionFull}), (\ref{ResonantConditionCycl}), and (\ref{ResonantConditionFullWhi}), led to the conclusion that inter-branch interactions are not allowed. The intra-branch processes (for both branches) are allowed and constrained by exactly the same relationship,  
\begin{equation}\label {krelation}
\frac{k_{2z}}{k_{1z}}=\frac {k_1-K}{k_2-K}\quad \Rightarrow \quad {k_{2z}}=\frac {k_1-K}{k_2-K}k_{1z}\equiv \delta(k)k_{1z}
\end{equation}
between $k_{1z}$ and $k_{2z}$. Thus in the low $k$ regime, showcased in the analytic model, the rest of this calculation does not distinguish  between the AW and AC branches. This expression could be used to eliminate either of the $k_z$ in favor of the other

Perhaps an additional comment may be useful. This analysis pertains to the slow scale modulation of energy of the three resonantly interacting waves; no claim is made that this attempt covers all the possible processes that constitute a general nonlinear Alfv\'enic state. Thus the demonstrated absence of the inter-branch interaction towards this particular end, does not rule out its possible importance to Alf\'venic turbulence.

For further progress, we must grapple with the immense $k$ variation that is embedded in the $J$ functions. One should remember that this variation reflects the complex nature of the Alf\'venic nonlinearities. Although three different  functions $[J_{12}, J_{2h}, J_{-1h}]$ are needed to map out the complete behavior , we will concentrate on explicitly discussing $J_{12}$; others have qualitatively similar features. This much of detail will be enough to establish basic scalings for  $t_{depletion}$. 

The next subsection, based on the material developed in Secs. IIIA-IIIB and Appendix.2, will examine how the interaction coefficient $J_{12}$ changes as the interaction wanders in the $[{\bf{k_1}}, {\bf{k_2}}]$ space. The small $ |k_z|/k$ assumption helps us avoid algebraic clutter. 
Three distinct representative regions, defined by the relative orientations of the perpendicular components of the participating vectors [the Zero, the $\pi$ and the $\pi/2$ sectors], are discussed in detail.  In all three subsections, the goal is to obtain approximate expressions, in fact, scalings for the respective depletion times.

\subsection{ The Zero Sector - $J^0$ and $t^{0}_{depletion}$}

In the zero sector, ${\bf k_{1{\perp}}}, {\bf k_{2{\perp}}}$ are near parallel, i.e, $(\theta_1-\theta_2) \equiv \Delta \theta <<1$ implying $\cos (\theta_1-\theta_2)\approx 1- (\Delta \theta)^2/2 , \sin (\theta_1-\theta_2)\approx \Delta \theta $. In addition $K\approx k_2-k_1 \Rightarrow K-k_2\approx -k_1, K-k_1\approx k_2 - 2k_1$

It must be emphasized, however, that parallel ${\bf k}_{{1\perp}}, {\bf k}_{2{\perp}}$ is not equivalent to the collinearity of ${\bf k}_1$ and ${\bf k}_2 $ for which case the interaction is identically zero; the latter will require the additional condition $ k_{1z}k_{2{\perp}}=  k_{2z}k_{1{\perp}}$.

The detailed working out of $t^{0}_{depletion}$ must be preceded by the following  observations: Since $K-k_1\approx k_2-2k_1$ and $K-k_2\approx-k_1$, two regions, leading to qualitative different asymptotic states, must be distinguished ($k_2>k_1$)

1) If  $k_2-2k_1<0 $, both $\sigma(\mu_1)$ and $\sigma(\mu_2)$ are positive (same as $\sigma(\mu_3)$), the evolution triplet 
looks like (\ref{pf+++}) with no directional energy exchange processes

2) For $k_2-2k_1>0 $ implying $\sigma(\mu_1)=+$ and $\sigma(\mu_2)= -$, the triplet evolution follows (\ref{pf+-+}); energy exchange processes transferring energy to the highest $k (namely k_2)$ will take place.    

Having settled that, we find that in the zero sector, the function
\begin{equation}
(K+k_1+k_2)^{3/2}[\frac{|K-k_1| |K-k_2| |k_2-k_1|}{K k_1 k_2}]^{1/2}\approx(2 k_2)^{3/2}|1-2q|)^{1/2}
\end{equation}
where $q=k_1/k_2<1$. If the $J$ dependence is further approximated as $(J^0)^{3/2}$, then using  Eq.(\ref{Jcyl0A}), we obtain
\begin{equation}\label{defining F}
 \frac{1}{ t^{0}_{depletion}}\sim   (\sqrt{2} k_2 k_{1z})^{3/2} \left(\frac{(1-q)^2}{(1-q)^2 + \epsilon^2}\right)^{3/2} |1-2q|^{1/2}\equiv  (\sqrt{2} k_2 k_{1z})^{3/2} F(q). 
\end{equation}
where $F(q)$, a function primarily of $q$ ($0<q<1$), has a zero at $q=1/2$. For $q< (>)1/2$, the energy exchange processes do (do not) occur. Of course, when such processes do occur, the energy will be shifted to the highest $k (=k_2)$ mode. 

In Fig.6, one sees that $F(q)$ has considerable structure; in the small q region, $F(q)$ decreases (monotonically )from 1 to zero, but in the $q>1/2$ region, it goes through a maximum before hitting zero at $q=1$. Barring the region near $q=1/2$, $F(q)$ may be approximated by a number less than unity. This leads to the final scaling
\begin{equation}\label{Dep0}
\frac{1}{ t^{0}_{depletion}}\sim ( k_2 k_{1z})^{3/2}; 
\end{equation}
the energy exchange processes go faster as $k_z$ as well as the modulus of the wave number go up. One must bear in mind, however, that  ${ t^{0}_{depletion}}\rightarrow\infty$ at $k_2=2k_1$.

\begin{figure}
\includegraphics[width=4in]{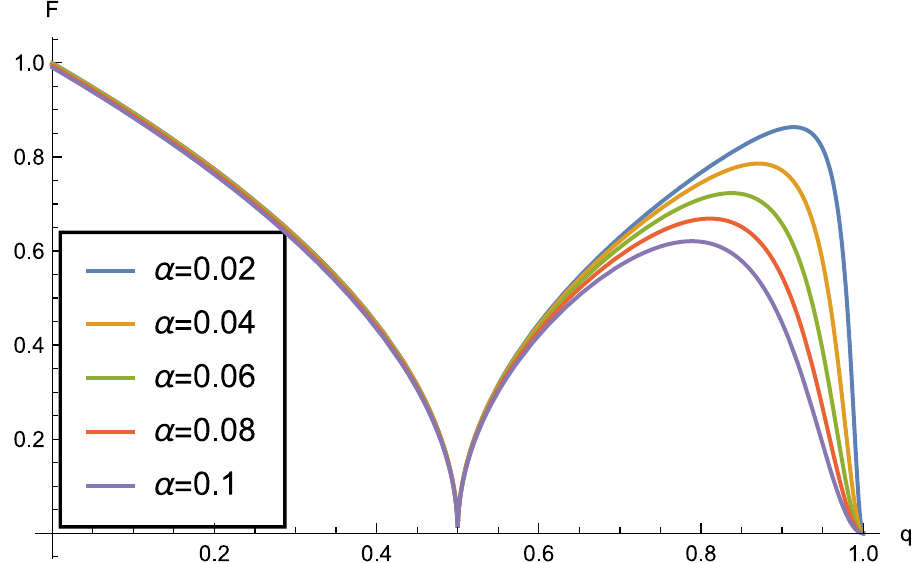}
\caption{ An illustration of the somewhat complicated k dependence of a typical interaction Kernel for nearly parallel (zero sector) wave vectors.  The function $F(q)$ (\ref{defining F}) is plotted against $q (0<q<1)$ for a set of  $\epsilon=a$ values.  The region $(0<q<1/2)$, where energy exchange processes are possible, goes monotonically to zero at $q=1/2$. The region where such processes do not occur, has lot more structure near  $q=1$. The peak can be quite accurately calculated. It would appear that in this part of the $k$-space, interactions are stronger as $k_1$ and $k_2 $ become further apart.} 
\end{figure}

\begin{figure}
\includegraphics[width=4in]{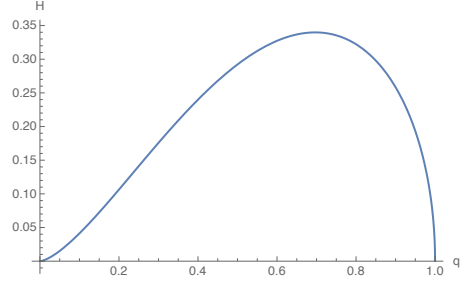}
\caption {H(q) plotted as function of q; it tends to peak around q=.75 before vanishing at q=1}
\end{figure}

\subsection{ The $\pi$ Sector} 

The $k$ numbers in the $\pi$ Sector, where the perpendicular wave vectors are nearly antiparallel(and $K\approx k_1+k_2$), belong to the  hierarchy $K>k_2>k_1$ - the hybrid $K$ is the largest wave number. The resulting set [$\sigma (\mu_1)=-1=\sigma(\mu_2), \sigma (\mu_3)=1$] puts this combination in the class of Eq. (\ref{pf--+}). Directional energy exchange processes that will eventually shift the initial energy into the highest K (hybrid) mode are allowed in this region of k-space. 

Following the pattern of the Zero sector analysis, we first evaluate
\begin{equation}
 (K+k_1+k_2)^{3/2}\left[\frac{|K-k_1| |K-k_2| |k_2-k_1|}{K k_1 k_2}\right]^{1/2}\approx(2 k_2)^{3/2} (1+q)^{-1/2} (1-q)^{1/2},  
\end{equation}
which in conjunction with Eq.(\ref{cylpi}) of Appendix 2, yields 
\begin{equation}
 \frac{1}{ t^{\pi}_{depletion}}\sim  4\left(\frac{k^{3}_{1z}}{k_2}\right)^{3/2} G(q),  \quad    G(q)= \left(\frac{1-q)}{1+q}\right)^{1/2}\left(\frac {1+q^2} {q^2}\right)^{3/2} 
\end{equation}
The function G(q) has two noteworthy features:  1) It tends to become larger when $q$ becomes small, and 2) it goes slowly to zero when $q\rightarrow 1$, i.e, when $k_1$ approaches $k_2$. Since we have assumed that $(k_{1z}<< k_1, k_2)$, q cannot become too small to compensate for the small factor $(k_{1z}/k_2)^2$. In most of the range, we can approximate  
 \begin{equation}\label{Deppi}
\frac{1}{ t^{\pi}_{depletion}}\sim  (\frac{k^{3}_{1z}}{k_2})^{3/2} \bar G ,  
\end{equation}
where 
$\bar G$ is a number of order unity except, perhaps, when $k_1$ is a small fraction of $k_2$. 

%

\subsection{ The ${\pi/2}$ Sector} 
The $\pi/2$ sector, where the perpendicular wave vectors are  nearly perpendicular to one another  (with $K\approx\sqrt {k^{2}_1+k^{2}_2}$), also belongs to the hierarchy $K>k_2>k_1$; the corresponding triplet dynamics is, thus, governed  by  (\ref{pf--+}).  Just like the $\pi$ sector, this region of k space will allow directional energy exchange processes that will, eventually, shift the initial energy into the highest $ k (= K)$ fluctuation. We start again with finding the appropriate form for
\begin{equation}
 (K+k_1+k_2)^{3/2}\left[\frac{|K-k_1| |K-k_2| |k_2-k_1|}{K k_1 k_2}\right]^{1/2}\approx( k_2)^{3/2}H(q). 
\end{equation}
\begin{equation}
 H(q)= (\sqrt{1+q^2} +1+q)^{3/2} \left[\frac {(\sqrt{1+q^2} -1)(\sqrt{1+q^2} -q)}{q\sqrt{1+q^2}}\right]^{1/2}(1-q)^{1/2} .
\end{equation}
At both ends of the q range, the function $H(q)\rightarrow$ 0 . Its peak value is of order unity (it is $\sim .35$ to be exact - see Fig.7).  Thus, keeping with the spirit of this analytic investigation, we find, after combining with $J_{\pi/2}$, given by (\ref{cylpiby2}) from Appendix 2, we have 
\begin{equation}\label{Deppi2}
\frac{1}{ t^{\pi/2}_{depletion}}\sim  k^{3}_{2},  
\end{equation}

\begin{table}[h]
    \centering
    \begin{tabular}{|c|c|c|c|}
         \hline
         \makecell{Sector $\rightarrow$ \\ Nature of coupling $\downarrow$} & 0 & $\pi$ & $\pi/2$ \\ \hline
         \makecell{Intra-branch\\ +++,$---$} & Moderate$\sim (k_2 k_{z})^{3/2}$ & Weak$\sim k_{z} (k^2_{z}/k)^2$ & Strongest$\sim k^{3}_2$  \\ \hline
          \end{tabular}
    \caption[font=5]{Interaction Strength }
    \label{T1}
\end{table}

\vspace{.4cm}

One cannot but notice the enormous difference in the the estimated depletion times in the three sectors. Taking  $t^{0}_{depletion}$
as fiducial reference (chosen arbitrarily), we find the ratios
\begin{equation}
R^{\pi}_0=\frac{ t^{\pi}_{depletion}}{ t^{0}_{depletion}}\sim (\frac{k_{2}}{k_{1z}})^3, \quad R^{\pi/2}_0=\frac{ t^{\pi/2}_{depletion}}{ t^{0}_{depletion}}\sim (\frac{k_{1z}}{k_2})^{3/2}:
\end{equation}
the nonlinear processes act enormously fast in the $\pi/2$ sector as compared to the zero sector, which, in turn, modulates the triplet wave interaction way faster than the $\pi$ sector: $[t^{\pi/2}_d <<  t^{0}_d << t^{\pi}_d]$. The three sector $k$ scaling is displayed in Table1.
\section {Perspective, Summary and  Discussion}
After this detailed presentation of  both the essence and technical aspects of the calculation, we are now ready to examine what we have learnt and what remains to be learnt within the current framework of nonlinear HMHD/extended MHD. 

I would like to begin the critical survey by stating that this paper, devoted to developing a conceptual framework for investigating nonlinear Alfv\'enic states (including turbulence), is, hopefully, the first of many. Although MHD turbulence has been studied rather extensively, there have been relatively few systematic investigations of fully nonlinear Hall/Extended MHD. Using HMHD as a representative example, we have developed a methodology that is applicable to most Extended MHD systems.

The transition from MHD to HMHD is rather nontrivial; in addition to capturing physics more accurately at shorter wave lengths, it acts to bring in at least two fundamental changes: 1) the system has a new Hall-induced nonlinearity in the Ohm's law, and 2) the character of the linear modes (waves) of the system undergoes quite a drastic change; the HMHD dispersion relation is satisfied only for circularly polarized vector fields (${\bf b}, {\bf v}$), and unlike simple MHD, modes with arbitrary polarization are no longer the normal modes of the system. One should also stress that the spectrum splits into two  branches - the Alfven-Whistler ( AW +) and the Alfven -Cyclotron (AC ) waves with distinctly different dispersion relations as we approach the moderate-to-large $k$ limit.  Both of these changes conspire to impart nonlinear HMHD a great deal more complexity, and substantial new content. 

Fortunately, the resulting constraint of circular polarization can be readily dealt with by resorting to a readily available alternative to vector Fourier transforms - the Beltrami transform. Extensively used previously in the study of linear as well as Linear-Nonlinear HMHD waves, the Beltrami transform allows a decomposition in terms of the Beltrami vectors. As they are eigenvectors for the curl operator, the new basis is ideally suited to handle HMHD nonlinearities that are so curl dominated. The resulting nonlinear formalism (in which all curl operators become scalar multipliers) consists of scalar equations that appear to be simple enough that an easier exploration and analysis by standard tools may be anticipated. Numerical simulations are likely to  be (much) less time consuming. Whether it turns out to be actually so, remains a question for future research, and is being seriously pursued concurrently (Private Communication: David Hatch).

The principal short term objective of this paper, however, was to investigate the nature and expression of the nonlinearities by carrying out a straightforward analysis made possible by the relative simplicity of the Beltrami-transformed HMHD equations. To keep the analysis focused and accessible, a model problem- the resonant interacting of a wave triplet  was worked out in detail to extract and expose the character of nonlinear interactions that control how, how fast and in what direction (in the wave vector space) do the energy exchanges take place. While this model calculation (limited to the regime $k_z<<k$) is admittedly not expected to simulate all essentials of all all turbulent states, it does nonetheless furnish crucial insights into the processes that will lead to one.

Perhaps, a supreme example of what makes the new formalism easier to explore is that much of the vectorial complications of nonlinearities is subsumed in the form of a single scalar Interaction Kernel $I_{mn}$. The kernel is, therefore, an encapsulation of the geometry of the ${\bf k}$ space. The total measure of the strength of the resonant interaction, however, comes from a product of $I_{12}$ and the factors $g_{nm}$ and $ h_{nm}$ (see Eq. \ref{gandf}); the latter reflect the interplay between the contributions from the three nonlinearities (${\bf v}\times {\bf b}, {\bf v}\times {\bf \nabla\times v}$, and ${\bf b}\times {\bf \nabla\times b}$); the interaction kernel is a geometric factor common to all three. The final manifestation of the ``nonlinearity" comes about through an intricate mix of its various constituents, and is a strongly sensitive function of the involved wave vectors.

It is the factors $ g_{nm}$ and $f_{nm}$ that determine what will be the nature of the resonant interaction as we comb the space spanned by the vectors [${\bf k}_1, {\bf k}_2$]. The triplet of evolution equations can be reduced to an extremely simple looking set  whose collective solution is dependent on whether the three $\sigma(\mu_i)$, defined in (\ref{Mu1})-(\ref{Muh}), and constructed from $[g, f]$, are positive or negative.  The signs, in turn, are determined by the relative ordering of the wave numbers $[K, k_1, k_2]$. If all $\sigma (\mu)$ have the same sign, the modes interact but without any effective energy transfer. If two of the $\sigma (\mu)$, however, are different from the third, energy exchanges between the modes constitutes the dominant dynamics; both periodic and directional transfers (in k space) are possible. For the latter, the direction of transfer is uniquely one way- always from the lower to the higher $k$. Asymptotically, all initial energy is transferred to the highest  $k$ in the triplet. The rates of energy exchange differ widely; for the readers convenience, approximate strength of the interaction, in several representative regions, is displayed in Table1.

Despite the abundance of specific technical results, the overarching message of this initial analytical effort is that the action of the nonlinear interactions are rather variegated. Consequently, the identification of a simple ``universal" description or a simple formula is not readily feasible. The class of general statements in the literature on MHD turbulence \cite{Bis03,Dav,BC16}: 1) only antiparallel modes can have strong coupling, 2) three wave couplings are not allowed, 3 ) in the allowed three wave couplings, one must be a zero frequency mode: are either directly contradicted by the current analysis or found to have a very limited validity in some region of $k$-space. Hence, elevating them to a level of generality suited for constructing a theory is premature. 

The particular results that impose this sobering conclusion are restated both for clarity and emphasis:  

\item There is no interaction between modes with collinear wave vectors. In particular, when $\hat { k_m}$ and $\hat {k_n}$ are either parallel or antiparallel, the Interaction Kernel $I_{mn}$ becomes identically zero, irrespective of the other parameters in the system. 

\item the Interaction Kernel $I_{nm}$, pertinent to the resonant interaction, varies greatly with the relative orientation of the ${\bf k_{n\perp}}$ and ${\bf k_{m\perp}}$, i.e, with the angle $(\theta_n-\theta_m)\equiv \Delta\theta$. This striking variation was, in fact, what prompted the convenient book-keeping nomenclature: the zero ($\Delta\theta\approx 0$), the  $\pi (\Delta\theta\approx \pi$) and the $\pi/2 (\Delta\theta\approx \pi/2)$ sectors.  

\item The geometrical effects that make  $(1+ \hat {\bf k}_n\cdot\hat {\bf k}_m) \approx (k_z/k)^2 <<1$, strongly suppress the effective strength of the nonlinearities in the $\pi$ sector (the two ${\bf k}_{\perp}s$ are nearly antiparallel). Conversely, the term 
$\hat e_z\cdot(\hat k_1 \times \hat k_2)$ that reaches its maximum when ${\bf k}_{n{\perp}}$ and ${\bf k}_{m{\perp}}$ are perpendicular, boosts the interaction upwards in the $\pi/2$ sector. As the relative wave vector orientation wanders from $0$ to $\pi$ in the plane perpendicular to the ambient magnetic field, the corresponding modes experience moderate to strong to quite weak   
coupling. 

\item It would appear that much of the structure of $I_{nm}$ is controlled by ${\bf k_{n\perp}}$ and ${\bf k_{m\perp}}$, by their relative orientation (qualitative) as well as their magnitudes (quantitative). It is, perhaps, expected that the curl  nonlinearities will valorize the perpendicular wave vector. 

\item In our model problem, we have introduced the idea of the depletion time $t_{depletion}$ as a measure of the overall strength/efficacy of the nonlinear interaction; $t_{depletion}$ is the time needed for a substantial inter-mode energy transfer. The fact that we have arrived at three strikingly different formulas for $t_{depletion}$ [Eqs. (\ref{Dep0}), (\ref{Deppi}), (\ref{Deppi2})] is telling.  The widely different $q=k_1/k_2$ dependence (represented, respectively by the factors F(q), G(q) and H(q)) for each of these is a further reminder that HMHD nonlinearities are quite complex.

\item Despite the variety in the exact forms of energy depletion/oscillation times time, there  are noticeable common features: 1) In the zero and $\pi/2$ sectors,  $t_{depletion}$ goes down rapidly with the magnitude of $k$, 2) barring the $(\pi/2)$ sector, the other two sectors show an additional $k_z$ dependence; the depletion times scale moderately to strongly with $k_z$; larger $k_z$ modes transfer energy faster to nearby higher k modes. Perhaps, the most universal signature of the nonlinearities, revealed in the analytical model, concerns the direction of  energy transfer; the energy always flows from the lower to the highest $k$ in the triplet enriching the high $k$ part of the spectrum. This cascade is consistent with prior numerical simulations and theoretical models of MHD and extended MHD turbulence \cite{Bis03}.

\item Pursuing the $k_z$ dependence is of essence. Our equations represent the interaction between three 'modes': $  [n](k_{nz}, k_{n\perp},\theta_n)$, $ [m](k_{mz}, k_{m\perp}, \theta_m)$ and the  hybrid [h=m-n] whose ${\bf K}$ vector is constructed from those of [n] and [m].  In effect, they represent interaction between two independent 'classes of 'modes' labelled by $n$ and $m$. Since $k_{nz}<<k_{n\perp}, k_n$, for a given  $k_n$, there can be a whole spectrum of ($k_{nz}, n=n_1, n_2--$). In this spectrum, the the modes with a large parallel wave number $k_{n_{1}z}(>k_{n_{2}2})$ will be depleted faster,  and if one were to wait long enough, the spectrum will become richer in low $k_z$. In a turbulent Alfv\'enic state, one could imagine a test mode $[n]$ undergoing a spectral readjustment interacting with a reference mode with somewhat higher wave number - the set of all test modes, forming the eventual turbulent state will be, preferentially, found in the lower part of the $k_z$ spectrum. Since the energy is also transferred upwards in $k$, the ratio of $k_z/k$ in the spectrum will continuously go down.  

\item The model problem solved in this paper ended up excluding the inter-branch coupling because the resonant condition $(\omega_n=\omega_m+\omega_{n-m})$ was hard to satisfy. One  must realize, however, all nonlinear processes in HMHD are not resonant ( the mode could scatter off  another, for instance); inter-branch interactions could and likely would play some part in the nonlinear evolution of an Alf\'venic system and merit further study. 

\item If one were to indulge in developing an argument analogous to critical balance by asserting that in the steady state turbulence the linear $(\omega^{-1}\approx k_z^{-1})$ and nonlinear time ($t_{depletion}$) approach one another, we will not get any unique answers given that ($t_{depletion}$) is subject to considerable variation. It is prudent to bypass such an exercise until we gain a deeper understanding of the system. 

\item A caveat is in order; in this first conceptual paper investigating the formalism described by the Beltrami-decomposed Hall MHD, I have not explored the region of $k$ space where $k_z$ and $k_\perp$ may be comparable. If this regime were to pertain for any given experiment or observations, such an analysis can be readily performed. In fact, in the explored part of $k$-space, we have essentially concentrated on the low $k$ MHD like regime. A more in-depth investigation of the whistler part of the spectrum is on the agenda.

\item In principle, the energy exchange mechanisms explored in this work can be directly tested  in laboratory experiments. One can design an experiment that replicates exactly the theoretical scenario:  Excite three  Alfvenic modes initially (conceivably with near equal initial amplitudes), and then observe the changes in energy distribution among them  as a function of time. 

\item Finally, I would like to make a comment on the possible role the conserved helicities would play in the context of the three wave resonant solutions we have explored in this paper ( A detailed analysis of the system helicities will be a major addition). The simplest role that the helicities can assume is to "define" the initial data for the initial value problem. If $h_1$ and $h_2$ were, respectively, the magnetic and the generalized helicities, then the initial energies $E$ and $aE$ can readily be related to them. In that sense, once the helicities are specified, the mode of evolution is essentially fixed  (apart from detailed k structure).

 The analysis of a simple model founded on a sufficiently well-posed and well-defined problem has revealed that  energy transfer processes in HMHD are sufficiently complicated that dwelling on one or the other nonlinear process may lead to inadequate and possibly erroneous conclusions. It seems that all nonlinearities have to be situated \emph{a priori} on equal footing and detailed calculations (in the vein of this paper are) are, perhaps, necessary to describe a nonlinear Alfv\'enic state. We have taken a modest step in this direction by demonstrating (analytically and quite convincingly) that the slow evolution of resonantly interacting modes does, preferentially, deposit energy, simultaneously into the higher $k$ and the lower $k_z$ part of the spectrum. But the potential of this formalism, in which the HMHD system has been reduced to  a relatively simpler set of scalar equations (through a novel expansion in terms of a complete set of Beltrami functions), is vast, and is manifestly ready for investigation via further analysis and detailed  numerical simulations.

 \end{itemize}

\section {Appendix. 1}

In this Appendix, we detail the steps needed to derive Eqs.(\ref {BHMHD1}-\ref {BHMHD2}) by taking the Beltrami transform of Eqs.(\ref{HMHD1})-(\ref{HMHD2}). Multiplying the latter by ${\bf Q}^*_m$ and integrating over space, we find
\begin{equation}
\frac{\partial b_m}{\partial t} - i k_{mz} ( v_m - k_m b_m )= N1
\end{equation}
\begin{equation}
k_m \left [ \frac{\partial v_m}{\partial t} - i k_{mz} b_m \right ] =N2
\end{equation}
where 
\begin{equation}\label{Semi1}
N1=\int {d^3x} \\\ (\nabla \times [{\bf v} \times {\bf b}  - (\nabla \times {\bf b} ) \times {\bf b}]) \cdot{\bf Q}^*_m=  \int {d^3x}\\\ (NL1)\cdot{\bf Q}^*_m ,
\end{equation},
\begin{equation}\label{Semi2}
N2=\int {d^3x}\\\ (\nabla \times [ {\bf v} \times (\nabla \times {\bf v})  + (\nabla \times {\bf b} ) \times {\bf b}]) \cdot{\bf Q}^*_m= \int {d^3x}\\\ (NL2) \cdot {\bf Q}^*_m ,
\end{equation}
are the transforms of the  nonlinear terms.

Expanding $\bf v$ and $\bf b$ via (\ref{Beltexpand}), and using the characteristic  properties of the Beltrami vectors (enumerated in Sec.I), various components constituting  NL1 and NL2, become
\begin{equation}\label{1}
 \nabla \times [{\bf v} \times {\bf b}]= \sum_{n}\sum_{l} v_{n}\\ b_{l}\\ \nabla\times({\bf Q}_n\times{\bf Q}_l)
\end{equation}
\begin{equation}\label{2}
\nabla \times [{\bf b} \times \nabla\times {\bf b}]=\sum_{n}\sum_{l} b_{n} b_{l} k_l \nabla\times({\bf Q}_n\times{\bf Q}_l)
\end{equation}
\begin{equation}\label{3}
\nabla \times [{\bf v} \times \nabla\times {\bf v}]=\sum_{n}\sum_{l} v_{n} v_{l} k_l \nabla\times({\bf Q}_n\times{\bf Q}_l)
\end{equation}
Since  Qs are divergence free, the common factor to (\ref{1}-\ref{3}) becomes 
\begin{equation}\label{4}
\nabla\times({\bf Q}_n\times{\bf Q}_l)= ({\bf Q}_l\cdot\nabla) {\bf Q}_n-({\bf Q}_n\cdot\nabla) {\bf Q}_l
\end{equation}
The special properties of the Beltrami functions are, again, helpful in converting differential operators to multipliers,
\begin{equation}\label{5}
 ({\bf Q}_l\cdot\nabla) {\bf Q}_n=  (Q_{lx_i}\partial x_i){\bf Q}_n=i ({\bf k}_n\cdot {\bf Q}_l){\bf Q}_n,\quad ({\bf Q}_n\cdot\nabla) {\bf Q}_l=
i ({\bf k}_l\cdot {\bf Q}_n){\bf Q}_l
\end{equation}
Substituting (\ref{5}) into Eqs. (\ref{Semi1}-\ref{Semi1}), one carries out the space integral which yields $\delta_{l, m-n}$, and then use it to sum over the index $l$, we derive ( K is defined in Eq.(\ref{hybridKT}))
\begin{equation}\label{Fin1}
N1= i \sum_{n}[ v_n b_{m-n} - v_{m-n} b_n  + (K - k_n ) b_n b_{m-n} ] I_{nm},
\end{equation}
\begin{equation}\label{Fin2}
N2= i \sum_{n} ( v_n v_{m-n} - b_n b_{m-n} ) ( K - k_n  ) I_{nm}
\end{equation}
where all the  vectorial complications of the nonlinear interaction are subsumed in the rather succinct  kernel
\begin{equation}\label{kernel}
I_{n m}=({\bf k}_n\cdot\hat e_{m-n})(\hat e_n\cdot {\hat e_m}^*)
\end{equation}
The interaction kernel (\ref{kernel}) is explicitly worked out in Appendix 2.

\section {Appendix. 2}
In this section we will calculate $I_{nm}$ from its definition 
\begin{align}\label{inmgen}
I_{nm} = ({\bf k}_n  \cdot \hat{e}_{m-n} )(\hat{e}_n \cdot \hat{e}_m^*).
\end{align}
where 
\begin{equation} \label{polgen}
\hat e_n= \hat{e}_{n1}+i \hat{e}_{n2}= \frac{k_n}{\sqrt {2}k_{n\perp}}[\hat{e}_z \times \hat{k_n} +i\hat{k_n} \times (\hat{e}_z \times \hat{k_n})]
\end{equation}
 Let us calculate
 \begin{equation} \label{polproduct}
 \hat{e}_n \cdot \hat{e}_m^*= \frac{k_nk_m}{2k_{n\perp}k_{m\perp}}[\hat{e}_z \times \hat{k_n} +i\hat{k_n} \times (\hat{e}_z \times \hat{k_n})]\cdot
 [\hat{e}_z \times \hat{k_m} -i\hat{k_m} \times (\hat{e}_z \times \hat{k_m})]\equiv \frac{X+iY}{2}
 \end{equation}
 where
 \begin{equation} \label{polstep1}
X= \frac{k_nk_m}{k_{n\perp}k_{m\perp}}[(\hat{e}_z \times \hat{k_n})\cdot (\hat{e}_z \times \hat{k_m}) +(\hat{k_n} \times (\hat{e}_z \times \hat{k_n}))\cdot
\hat{k_m} \times (\hat{e}_z \times \hat{k_m})],
\end{equation}
\begin{equation} \label{polstep2}
Y= \frac{k_nk_m}{k_{n\perp}k_{m\perp}}[ (\hat{k_n} \times (\hat{e}_z \times \hat{k_n}))\cdot (\hat{e}_z \times \hat{k_m}) - \hat{k_m} \times (\hat{e}_z \times \hat{k_m})\cdot (\hat{e}_z \times \hat{k_n})]  
 \end{equation}
becoming
 \begin{equation} \label{polstep3}
X=(1+\hat k_m \cdot \hat{k_n}) \frac{{\bf k_{n\perp}} \cdot {\bf k_{m\perp}}}{k_{m\perp} k_{n\perp}}+ \frac{(k_{nx}k_{my}-k_{ny}k_{mx})^2}{k_n k_m k_{m\perp} k_{n\perp}}
\end{equation}
and 
 \begin{equation} \label{polstep4}
Y=(\hat{k_n}+\hat{k_m})\cdot\hat e_z \frac{(k_{nx}k_{my}-k_{ny}k_{mx})}{k_{m\perp} k_{n\perp}}.
\end{equation}
 In deriving these equations we have explicitly used the general definition of ${\bf k}$
\begin{equation}\label{kgen}
{ \bf k}=k_z \hat e_z +k_x \hat e_x+k_y \hat e_y\equiv= k_z \hat e_z +{\bf k_{\perp}}
\end{equation}
 To evaluate the second factor 
\begin{equation}\label{Kgen}
{\bf k}_n  \cdot \hat{e}_{m-n}= \frac{K}{\sqrt {2}K_{\perp}} {\bf k}_n  \cdot [\hat{e}_z \times \hat{K} +i\hat{K} \times (\hat{e}_z \times \hat{K})]\equiv \frac{G+iH}{\sqrt2},
\end{equation}
we need to define $\hat{K}={\bf K}/{K}$ with ${\bf K}\equiv{\bf K}_{m-n}={\bf k}_m -{\bf k}_n, K=|{\bf K}|$. Straightforward algebra leads to 
\begin{equation}\label{hybridK6}
 G= \frac{k_{mx}k_{ny}-k_{nx}k_{my}}{K_{\perp}}, \quad  H=\frac{[{\bf k_{m\perp}}k_{nz}-{\bf k_{n\perp}}k_{mz}]\cdot{\bf K}_{\perp}}{KK_{\perp}}
\end{equation}
One cannot help but notice that the second term of X, and the entire Y and G are all proportional to z component of $({\bf k_n} \times {\bf k_m})$, that is, $\hat{e}_z\cdot ({\bf k_n} \times {\bf k_m}) \equiv (k_{nx}k_{my}-k_{ny}k_{mx})$. Further, X and H are even while Y and G are odd on the exchange $nm<->mn$.We can now put together   
\begin{equation}
 i I_{nm}= \frac{1}{2\sqrt2} [-(XH+YG)+i(XG -YH)]\equiv [-A_{nm} +iB_{nm}]= J_{nm} e^{-i\zeta_{nm}}
\end{equation}
alongwith
\begin{equation}\label{IPhase}
 i I_{mn}=[-A_{nm} - iB_{nm}]=J_{nm} e^{i\zeta_{nm}}
\end{equation}
where $J^2_{nm}=A_{nm}^2+B_{nm}^2$ and $\zeta_{nm}=\arctan(B_{nm}/A_{nm})$; $I_{mn}$ has the same magnitude but opposite phase to  $I_{mn}$ 

\subsection {Interaction kernel in cylindrical k represntation} 
Substituting the cylindrical decomposition of ${\bf k}$ in X,Y, G and H, we find
 \begin{equation} \label{Xcyl}
X=[1+\frac{k_{nz}k_{mz}}{k_n k_m}+  \frac{k_{n{\perp}}k_{m{\perp}}}{k_n k_m} \cos (\theta_m-\theta_n)]  \cos (\theta_m-\theta_n)+ \sin^{2}(\theta_m-\theta_n)        
\end{equation}
\begin{equation} \label{YGcyl}
Y=(\frac{{k_{nz}}}{k_n}+\frac{{k_{mz}}}{k_m})\sin (\theta_m-\theta_n),  \quad  G=- \frac{k_{1\perp}k_{2\perp}\sin (\theta_m-\theta_n)}{K_{\perp}}
\end{equation}\label{Hcyl}
and 
\begin{equation}
H=\frac{[k_{nz}(k^2_{m{\perp}}-k_{m{\perp}}k_{n{\perp}}\cos (\theta_m-\theta_n))-k_{mz}(-k_n^2{\perp}+k_{m{\perp}}k_{n{\perp}}\cos (\theta_m-\theta_n))]}{KK_{\perp}}
\end{equation}
and 
\begin{equation}
K_{m-n}^2=k_{m}^2 + k^2_{n} -2k_{n{\perp}}k_{m{\perp}}\cos (\theta_m-\theta_n)
\end{equation}
We will need the values of $K_{12}=K$ in all three sectors ($(\theta_m-\theta_n)\approx 0, \pi, \pi/2$)
\begin{equation}\label{K0}
K_0^2\approx (k_{2z} - k_{1z})^2+ k_{1{\perp}}k_{2{\perp}}(\Delta \theta)^2 +(k_{2{\perp}}-k_{1{\perp}})^2 
\end{equation}
\begin{equation}\label{KpiK2}
K^2_{\pi}\approx  (k_{2{\perp}}+k_{1{\perp}})^2, \quad \quad K^2_{\pi/2}\approx  k_{2{\perp}}^2+k_{1{\perp}}^2
\end{equation}
We will treat $k_{\perp}\approx k$ wherever it is accurate.

We will now all set to evaluate the Interaction Kernel in all three limits:
\begin{itemize}
\item $J$ in the Zero Sector: In this sector $\bf{k}_{1,\perp}$ and $\bf{k}_{2,\perp}$ are near parallel, i.e, $(\theta_1-\theta_2) \equiv \Delta \theta <<1$ implying $\cos (\theta_1-\theta_2)\approx 1- (\Delta \theta)^2/2 , \sin (\theta_1-\theta_2)\approx \Delta \theta $. We find (after putting  $k\approx k_{\perp}$ in all noncritical places)
\begin{equation} \label{Xcyl0}
X\approx 2, \quad Y=0, \quad G=0, \quad H= \frac {(k_{1z}- q k_{2z})(1-q)}{(1-q)^2 + \frac { (k_ {1z}-k_{ 2z})^2}{k^2_2}}
\end{equation}
where 
$$q =\frac {k_{1\perp}} { k_{1\perp}} \approx \frac {k_1}{ k_2} <1$$
is the ratio of the interacting wave numbers, and in $K^2_0$, we have kept the $k_z$ dependent term so that there is a finite limit when $q$ approaches $1$. The corresponding Interaction Kernel  $J$ (Eq.\ref{J}) reduces to
$$J^0 = - \frac{XH}{2\sqrt 2}\approx= -\frac{1}{\sqrt 2} H^0; $$
it is $H^0$ that controls the rather sensitive $k$ behavior of the Interaction. Let us explore it further. Since the resonant condition $\omega_2=\omega_1+\omega_h$ is equivalent to $k_{2z}=\delta(k)k_{1z}$ where $\delta(k)= K-k_1/K-k_2\approx2-k_2/k_1=2-1/q$. We can, then, write
\begin{equation} \label{Jcyl0A}
J^0\approx -\frac {k_{1z}}{\sqrt2} \frac {(1-q)^2}{(1-q)^2 + \epsilon^2}, \quad \epsilon ^2\approx \frac{k^2_{1z}}{k^2_2}+q (\Delta \theta)^2.
\end{equation}

\item $J$ in the $\pi$ Sector: When ${\bf k}_{1{\perp}}, {\bf k}_{2{\perp}}$ are near antiparallel, i.e, $(\theta_1-\theta_2) = \pi+\Delta \theta (<<1)$, we approximate $\cos (\theta_1-\theta_2)\approx -1+(\Delta \theta)^2/2,  \sin (\theta_1-\theta_2)\approx -\Delta \theta $ leading to $(K_{\pi}\sim (k_1+k_2))$
\begin{equation} \label{cylpi}
Y\sim -(\frac{{k_{1z}}}{k_1}+\frac{{k_{2z}}}{k_2})\Delta \theta, \quad G\sim\frac{k}{2}\Delta \theta
 , \quad H\sim \frac{(k_{1z}k_{2\perp} + k_{2z}k_{1\perp})}{k_1 +k_2}
\end{equation}
In this sector, the resonance condition,  
\begin{equation} \label{resApp}
\frac{k_{2z}}{k_2}=\frac{k_{1z}}{k_1}
\end{equation}
will lead to
\begin{equation}
X\sim  2 (\frac{k_{1z}}{k_1})^2 +\frac{3}{2}(\Delta \theta)^2 \sim \epsilon^2, \quad H\sim \frac{k_{1z}(1+q^2)}{1 +q}
\end{equation}
and for $\Delta \theta=0$, the largest term contributing  to the interaction Kernel  
\begin{equation} \label{cylpi}
XH\sim -\frac{k^{3}_z}{k^{2}_2}  \frac {1+q^2}{(1 +q) q^2}\longrightarrow J_{\pi}\sim - s* k_z(\frac{k_z}{k_2})^2 \frac {1+q^2}{(1 +q) q^2}
\end{equation}
is two orders of magnitude smaller than the zero sector; s* is a number of order unity dependent on whether $k_{2z}$ is plus  (intra branch) or minus (inter-branch) $2k_{1z}$

\item $J$ in the $\pi/2$ Sector:  Because of the presence of the terms that are proportional to $[\hat e_z\cdot{\bf k_1} \times {\bf k_2})]$ in X,Y and G, the interaction Kernel can become of oder unity when ${\bf k_{1\perp}}$ and ${\bf k_{2\perp}}$
are perpendicular, i.e,  when $(\theta_1-\theta_2) = \pi/2$. Since leading order terms are large we will ignore small departures from            $\pi/2$. One finds, at exact perpendicularity $(K^{2}_{\frac{\pi}{2}}= k^2_1+ k^2_2)$,
\begin{equation}
X\rightarrow 1, \quad Y\rightarrow (\frac{{k_{1z}}}{k_1}+\frac{{k_{2z}}}{k_2}), \quad G=-\frac{k_{1\perp}k_{2\perp}}{K},
\quad H= \frac {k_{1z} k^{2}_{2\perp}-k_{2z} k^{2}_{1\perp}}{K^2};
\end{equation}
 the largest term XG is of order unity ( in $k_z/k$ )while all other combinations are $\sim (k_z/k)$ or smaller. Thus in the $\pi/2$ sector, $iI_{12}$ is dominantly imaginary, and leads to  
  \begin{equation}\label{cylpiby2}
J_{\frac{\pi}{2}}=\frac {1}{2\sqrt2}XG \approx - \frac{q k_{2}} {\sqrt {1+q^2}}, \quad \zeta_{12}=\frac{\pi}{2}
\end{equation}
Thus the field amplitudes $b_1$ and $b_2$ will have a phase difference of $\pi/2$; this phase difference , however, has no consequences for the energy exchange processes between the two fields.
\end{itemize}

\section{Appendix. 3- Deriving the truncated Modulated Alfv\'en Waves }

We will now get 
\begin{equation}\label{EqnEvolutionA}
\frac {d\psi_m}{dt}= -\frac {1}{\pm 2\sqrt{1+k_{m}^2/4} }\sum_{n}  iI_{nm}  [\alpha_m g_{nm} +  f_{nm}] \psi_{n} \psi_{m-n}
\end{equation}
ready for analysis. We first notice that apart from the phase associated with $iI_{nm}= J_ {nm}exp ( -i\zeta_nm)$ (\ref{IPhase}), all other coefficients in (\ref{EqnEvolutionA})are real. It is possible to extract a $k$ dependent  phase so that we may deal with real amplitudes. Substituting 

\begin{equation}\label{EqnEvolutionA}
  \psi_m =  \Psi_m e^{i\xi_m}
\end{equation}
and imposing the condition $\xi_m= \xi_n+\xi_{m-n}-\zeta_{nm}$ (a constraint in the wave number space), we will arrive at 
\begin{equation}\label{EqnEvolutionA}
\frac {d\Psi_m}{dt}= -\frac {1}{\pm 2\sqrt{1+k_{m}^2/4} }\sum_{n}  J_{nm} [\alpha_m g_{nm} +  f_{nm}] \Psi_{n} \Psi_{m-n}
\end{equation}
where all $\Psi$ are real. From the original reality condition $b_m=-b_m^*$,  it follows that $\Psi_{-m}=-\Psi_m$. 

By taking m=1, 2, h,  (\ref {EqnEvolution}) yields the relatively simple but complete system 
\begin{equation}\label{EqnEvolution1}
\frac {d\Psi_1}{dt}= \frac {J_{21}} {2}[\alpha_1 g_{21} +  f_{21}] \Psi_{2} \Psi_{h}
\end{equation}
\begin{equation}\label{EqnEvolution2}
\frac {d\Psi_2}{dt}= \frac {J_{12}}{2}[\alpha_2 g_{12} +  f_{12}] \Psi_{1} \Psi_{h}
\end{equation}
\begin{equation}\label{EqnEvolutionh}
\frac {d\Psi_h}{dt}= \frac { J_{2h}-  J_{-1h}}{2}[\alpha_h g_h +  f_h] \Psi_{1} \Psi_2
 \end{equation}
of three coupled, first order evolution equations. The expressions for $g$ and $f$ [see Eq.(\ref{gandf})] are $\alpha$  dependent implying that they are functions only of $k_1, k_2$ and $K=|{\bf k_2}-{\bf k_2}|$ but depend upon the dispersion relation while the interaction matrices $I$, though determined by the vectors ${\bf k_1}$ and ${\bf k_2}$, have no knowledge of the dispersion relation. In pursuit of our goal of a tractable analytical model, we will work out in detail, here,  one representative example: the low k MHD limit of the Alfv\'en- Whistler  branch. In this limit, simple algebra leads to 
\begin{equation}\label{Mu1A}
\alpha_1 g_{21} +  f_{21}=-\frac{K-k_2}{2k_1}[k_1+k_2+K],
\end{equation}
\begin{equation}\label{Mu2A}
\alpha_1 g_{12} +  f_{12}=-\frac{K-k_1}{2k_2}[k_1+k_2+K],
\end{equation}
\begin{equation}\label{MuhA}
\alpha_h g_{h} +  f_{h}=-\frac{k_1-k_2}{2K}[k_1+k_2+K]
\end{equation}
These equations are easy enough to solve exactly in terms of Jacobian Elliptic functions. 

\section {Appendix. 4}

The  general triplet evolution set
\begin{equation}\label{parameterfreeAp}
 \frac {d\Psi_1}{d\tau}= \sigma(\mu_1) \Psi_{2} \Psi_{h}, \quad  \frac {d\Psi_2}{d\tau}= \sigma(\mu_2)\Psi_{h} \Psi_{1},\quad \frac {d\Psi_h}{d\tau}= \sigma(\mu_3) \Psi_{1} \Psi_{2},
\end {equation}
will, now, be examined in some detail to supplement the content of Sec.V. We will, first, show that there are really two independent 
$\mu$ combinations, from which all others can be constructed. In what follows positive or negative refers to $\sigma(\mu_i)$.

\begin{itemize}
\item All are positive: 
\begin{equation}\label{parameterfreeAppos}
 \frac {d\Psi_1}{d\tau}= \Psi_{2} \Psi_{h}, \quad  \frac {d\Psi_2}{d\tau}= \Psi_{h} \Psi_{1},\quad \frac {d\Psi_h}{d\tau}= \Psi_{1} \Psi_{2},
\end {equation}
This totally symmetric system, goes to all negative if either one or three of the amplitudes changes sign , i.e, if $[\Psi_1,\Psi_2,\Psi_h]$ solves (\ref{parameterfreeAppos}) , then $[-\Psi_1,\Psi_2,\Psi_h]$ (with all variants) and $[-\Psi_1,-\Psi_2,-\Psi_h]$ will solve 
\begin{equation}\label{parameterfreeAppos}
 \frac {d\Psi_1}{d\tau}= -\Psi_{2} \Psi_{h}, \quad  \frac {d\Psi_2}{d\tau}= - \Psi_{h} \Psi_{1},\quad \frac {d\Psi_h}{d\tau}= -\Psi_{1} \Psi_{2},
\end {equation}
All these variations will have the same energies measured by the squared amplitudes. Thus it is enough to analyze (\ref{parameterfreeAp}) for this group.

\item two are $+(-)$and the third is $-(+)$. A typical member is 
\begin{equation}\label{--+Ap}
 \frac {d\Psi_1}{d\tau}= -\Psi_{2} \Psi_{h}, \quad  \frac {d\Psi_2}{d\tau}= - \Psi_{h} \Psi_{1},\quad \frac {d\Psi_h}{d\tau}= \Psi_{1} \Psi_{2},
\end {equation}
Let us call it $[--+]$. Again,  if $[\Psi_1,\Psi_2,\Psi_h]$ is the solution of  $[--+]$, then $[\Psi_1,\Psi_2, -\Psi_h]$ will saitsfy
$[++ -]$. All other combinations of this class can be constructed by analogy. Similarly, one goes from $[-- +]$ to  $[+-+ ]$ by the transformation $[\Psi_1,\Psi_2,\Psi_h]$ to $[\Psi_1,-\Psi_h,\Psi_2]$. The entire set of $\mu$ permutations can be, thus, covered by analyzing (\ref{pf--+}) and (\ref {pf+++}) of the main text. 

Equation (\ref{pf--+}), obeying the conservation laws ($E$, $a$ >0)
 \begin{equation}\label{Energy1-2}
\Psi_{1}^2+ \Psi_{h}^2=E, \quad  \Psi_{2}^2+ \Psi_{h}^2= aE \quad \Rightarrow \Psi_{1}= \sqrt{E - \Psi_{h}^2}, \quad  \Psi_{2}= \sqrt{aE - \Psi_{h}^2}  
 \end{equation}
was converted to  
 \begin{equation}\label{FinAmplitudeAP}
\frac{d\hat{\Psi_h}}{dT} =  \sqrt{1- \hat{\Psi_{h}}^2} \sqrt{a - \hat{\Psi_{h}}^2}
 \end{equation}
The solutions of (\ref{FinAmplitudeAP}) are bounded; the exact solution may be written as 
 \begin{equation}\label{EllipticBounded1}
T\pm\chi_h = F(\phi, a^{1/2}), \quad  \phi=\arcsin \frac{\hat{\Psi_h}}{a^{1/2}},   \quad a <1 
 \end{equation}
and 
 \begin{equation}\label{EllipticBounded2}
T\pm\chi_h = \frac{1}{F(\chi, a^{-1/2})}, \quad  \chi=\arcsin \hat{\Psi_h}, \quad a >1 
 \end{equation}
where F is the Elliptic integral. The form displayed in the main text are the limits $a \approx 1$ when both integrals approach the same value and $a<<1(a>>1)$ for the first (second).

The Elliptic integral formulation for the finite (interesting) version of (\ref {pf+++}) ( $\mathcal E= - \mathcal E$), 
  \begin{equation}\label{FinAmplitudeAP}
\frac{d\hat{\Psi_h}}{dT} =  \sqrt{ \hat{\Psi_{h}}^2-1} \sqrt{\hat{\Psi_{h}}^2-p}, 
 \end{equation}
allows the solution 
\begin{equation}\label{EllipticBounded3}
T\pm\chi_h = F(\phi, p^{1/2}), \quad  \phi=\arcsin (\frac{\hat{\Psi_h}^2 -1 }{\hat{\Psi_h}^2 -p })^{1/2},  \quad p <1 
\end{equation}
from which one could derive the results in the main text. However, for this case in particular, taking limits of the elliptic integral is lot more involved than the solution of the approximate differential equation.

\end{itemize}

\begin{acknowledgments}
I have benefitted greatly from discussion with, and help from many colleagues at IFS and outside. I owe special thanks to Manasvi Lingam, David Hatch, and Richard Hazeltine for their persistent inputs though out this long lasting endeavor.  Michael Halfmoon, Joel Larakers and Prashant Valanju  are thanked for helping with the preparation of the manuscript. The work of SMM was supported by USDOE Contract No.DE-- FG 03-96ER-54366.  
\end{acknowledgments}

\section{Data Availability}
The data that support the findings of this study are available from the corresponding author upon reasonable request.

\end{document}